\documentclass[aoas]{imsart}

\RequirePackage{amsthm,amsmath,amsfonts,amssymb}
\usepackage{bm}
\usepackage{mathrsfs}
\usepackage{subfig}
\usepackage{latexsym}
\usepackage{upgreek}
\usepackage{multirow}
\usepackage{verbatim}
\usepackage{amsfonts}
\usepackage{amsmath,amssymb,amsfonts}
\usepackage{algorithmic}
\usepackage{dirtytalk}
\usepackage{tabularx}
\usepackage{multirow}
\usepackage{makecell}
\usepackage{verbatim}
\usepackage{amsfonts}
\usepackage{amsmath,amssymb,amsfonts}
\usepackage{algorithmic}
\usepackage{dirtytalk}
\usepackage{tabularx}
\usepackage{multirow}
\usepackage{makecell}
\usepackage{placeins}
\usepackage{float}
\RequirePackage[authoryear]{natbib}
\RequirePackage[colorlinks,citecolor=blue,urlcolor=blue]{hyperref}
\RequirePackage{graphicx}

\startlocaldefs
\numberwithin{equation}{section}
\theoremstyle{plain}
\usepackage[english]{babel}

\endlocaldefs

\begin{document}

\begin{frontmatter}
\title{Robust Gaussian Process Regression with Huber Likelihood  \thanksref{T1}}
\runtitle{Robust Gaussian Process with Huber density}
\thankstext{T1}{Supported by the Department of Energy via the Lawrence Livermore National Laboratory.}

\begin{aug}
\author[A]{\fnms{Pooja}~\snm{Algikar}\ead[label=e1]{apooja19@vt.edu}},
\author[B]{\fnms{Lamine}~\snm{Mili}\ead[label=e2]{lmili@vt.edu}}
\address[A]{\printead[presep={Electrical and Computer Engineering,\ Virginia Tech }]{e1}}

\address[B]{\printead[presep={Electrical and Computer Engineering,\ Virginia Tech }]{e2}}
\end{aug}

\begin{abstract}
Gaussian process regression in its most simplified form assumes normal homoscedastic noise and utilizes analytically tractable mean and covariance functions of predictive posterior distribution using Gaussian conditioning. Its hyperparameters are estimated by maximizing the evidence, commonly known as type-II maximum likelihood estimation. Unfortunately,  Bayesian inference based on Gaussian likelihood is not robust to outliers, which are often present in the observational training data sets. To overcome this problem, we propose a robust process model in the Gaussian process framework with the likelihood of observed data expressed as the Huber probability distribution. The proposed model employs weights based on projection statistics to scale residuals and bound the influence of vertical outliers and bad leverage points on the latent functions' estimates while exhibiting a high statistical efficiency at the Gaussian and thick-tailed noise distributions. The proposed method is demonstrated by two real-world problems and two numerical examples using datasets with additive errors following thick-tailed distributions such as Student's t-, Laplace, and Cauchy distribution.
\end{abstract}

\end{frontmatter}


\section{INTRODUCTION}
\label{sec:intro}
Bayesian inference based on Gaussian likelihood is known to be sensitive to extreme observations and gross errors, called outliers. In practice, the predictive posterior distribution is greatly influenced by outliers, which commonly arise due to human errors, measurement errors, meter sensitivity to changing weather conditions, and systematic variation unconsidered in the simplified model, to name a few. The two types of outliers likely to occur in a linear regression model are vertical outliers and bad leverage points. A vertical outlier is an outlier in the response space whose projection on the input vector space associated with the input variables is an inlier. A leverage point is a data point whose projection on the input vector space is an outlier. A leverage point can be a good or a bad point depending on whether it is an inlier or an outlier in the response space (\cite{Huber2004}). 

\begin{figure}[t]%
\centering
\subfloat[\centering  ]{{\includegraphics[height=5cm,width=5.7cm]{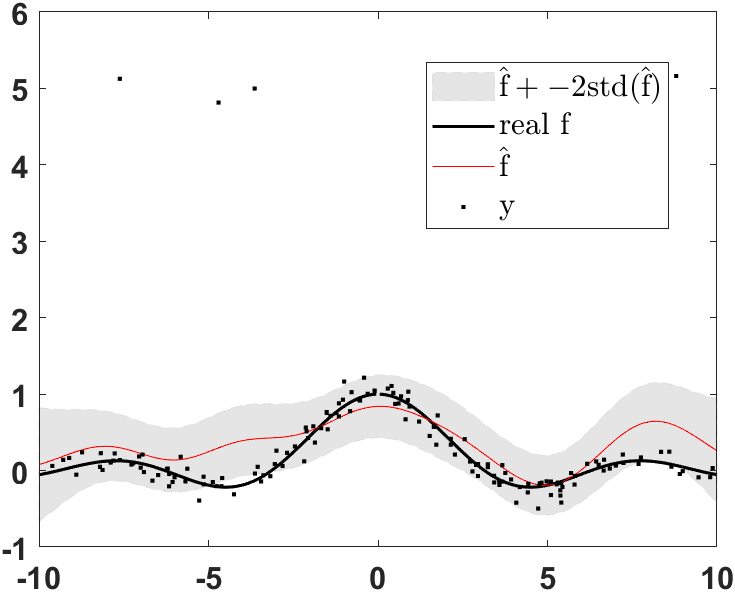}}}%
\quad \quad \quad 
\subfloat[\centering ]{{\includegraphics[height=5cm,width=5.7cm]{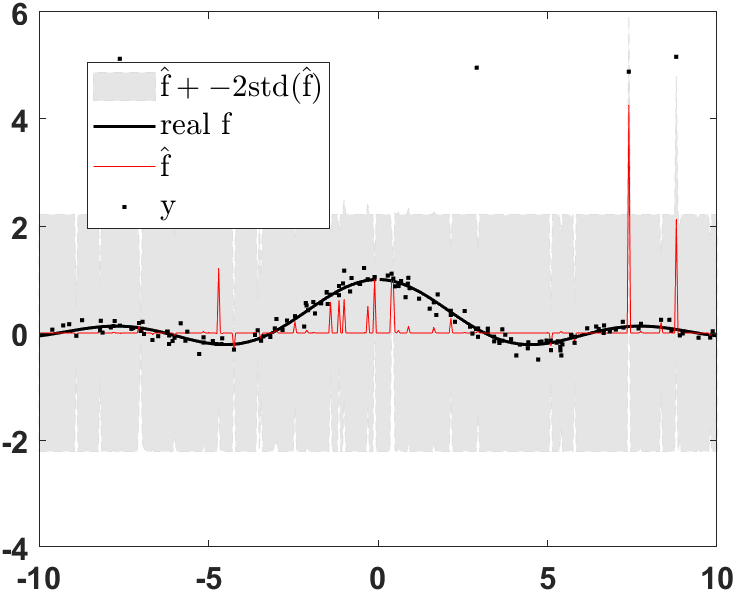} }}%
\caption{\textit{Results of the sinc function where the added errors are $i.i.d$ according to the Student-t distribution with $2$ degrees of freedom, that is, $e \sim \textrm{Student's t}(2)$, obtained from (a) GP with MAP estimates and (b) Student's t-likelihood with the MCMC integration approximation method.}}%
    \label{Laplace}%
\end{figure}

In Gaussian process regression models, the measurement error is assumed to be normal homoscedastic, resulting in biased parameter estimations of the posterior predictive distribution in a non-Gaussian noise setting. The predictive uncertainty attributes more confidence to the outliers than it should. Let us illustrate this problem in a numerical example. Consider a sinc function with an additive error that follows the Student's t-distribution with $2$ degrees of freedom. which is expressed as, $y(x)=\textrm{sinc}(x)+e$. Figure \ref{Laplace}(a) displays the predicted values at test points over the interval $[-10, 10]$ and spaced $0.01$, obtained from a Gaussian process (GP) regression. The uncertainty of the predicted values represented as their $\pm 2$ standard deviations is high.  The mean predicted values deviate largely from the true values of the sinc function. The caveat above is exacerbated by multiple outliers masking one another in the multivariate regression residuals.

Let us now consider a real-world example in the field of astronomy. In transmission spectroscopy of an exoplanet, the sources of noise such as photon noise, and instrumental and astrophysical systematics, raise many potential challenges for accurate estimation of planet-to-star radius ratio and precise atmospheric characterization. Pointing drift or modifications in telescope focus alter the detector spectrum position to a small degree as the planet passes in front of a star known as transit due to instrumental systematics. The fact that the measurements obtained from the HST-NICMOS instrument of an exoplanet, HD $189733$, are extremely contaminated has motivated us to develop a reliable robust GP-Huber method to perform the non-linear regression to infer the planet-to-star radius ratio of HD $189733$ planetary system. The estimates of the planet-to-star radius ratio are compared with that obtained using the state-of-the-art method developed by \cite{Gibson2011}.

A number of methods are proposed in the literature to robustify the GP regression models. For instance, the method proposed by 
\cite{Box1968} assumes that the measurement error follows a mixture of two normal distributions. Then \cite{West1984} investigated heavy-tailed error distributions that are constructed as scale mixtures of normal distributions, which are also used for specifying a prior distribution based on the earlier ideas suggested by  \cite{Finetti1961} and \cite{Ramsay1980}. By doing so, the a priori distribution discounts any observations highlighting inconsistency between likelihood and prior. Along the same line, \cite{Desgagné2019} assumed a super heavy-tailed error distribution dependent on an explanatory variable to make the estimation of the population mean and ratios robust to outliers. 
\cite{Kuss2006} extended a mixture of two normal distributions, one to model small errors in regular observations and a second one to model large errors in outlying observations. However, \cite{Naish2007} questioned the adequacy of the two-model approach. They proposed instead twin Gaussian processes that allow us to choose between the distribution of the regular observations and that of the outliers. \cite{Kuss2006} suggested a GP with a Laplace likelihood model that utilizes a scale mixture representation of Laplace noise distribution where the variance follows an exponential distribution. More recently, \cite{Vanhatalo2009} proposed a GP model based on the Student's t-likelihood function, where the noise is modeled as a scale mixture of Gaussian distributions. Unfortunately with the non-Gaussian likelihood, the Bayesian inference becomes analytically intractable. Consequently, various advanced approximation methods  were proposed (\cite{Kuss2006}, \cite{Vanhatalo2009}, \cite{Ranjan2016}) to overcome the convergence failure of the classical approximation methods such as expectation propagation (\cite{Minka2013}), Markov Chain Monte Carlo (\cite{Neal1997}), variational Bayes (\cite{Ghahramani2000}), and Laplace approximation (\cite{Williams1996}). For all these methods, except the one proposed by \cite{Vanhatalo2009}, the error distribution is required to be known a prior, which is not always possible in practical applications. Furthermore, when the large errors follow the same distribution as the likelihood of the observations, the robustness of the GP regression model is questionable. This is demonstrated in Figure \ref{Laplace}(b) using the sinc function for the GP with the Student's t-likelihood based on the Laplace approximation method. We notice that the model overfits while the uncertainty about the function is increased dramatically. 
 
In this paper, we address the problem of regression modeling under a wide class of heavy-tailed error distributions with smooth heteroscedasticity and strategically located large outliers. Our main contribution is the development of a unified robust regression model that is able to handle a broad range of error distributions,  including the Gaussian distribution, which may or may not be known a priori, and extreme outliers without principal changes in the likelihood function specific to the error distribution and without introducing any additional parameters pertinent to the likelihood function. Specifically, we propose a robust regression model in the GP framework based on a loss function well-established in robust statistics for observations that may have non-Gaussian responses, namely the Huber loss function, which relies on robust estimators of central trend and variability. The inference with a complex posterior distribution is implemented using the Laplace approximation and Markov Chain Monte Carlo (MCMC) methods. We show that the approach enables the parameter estimation to be  robust to outliers and attain maximum breakdown point. The task is accomplished by standardizing the residuals via  weights calculated using projection statistics applied to the data.  Introduced by \cite{Stahel} and \cite{Donoho1982}, the projection statistic of a multivariate data point relative to a multivariate point cloud is defined as the maximum of the absolute values of the standardized projections of that data point on the straight lines that pass through the coordinatewise median and  the data points (\cite{Rousseeuw2005}). It is a robust version of the Mahalanobis distance based on the sample median and the median absolute deviation from the median instead of the sample mean and the sample standard deviation. For illustrative purposes, we employ two numerical datasets, namely Neal and Friedman, with additional noise following a wide range of thick-tailed distributions. The predictive accuracy of the GP-Huber is compared with the Gaussian process regression models proposed with the Student's t- and Laplace likelihood.

\par The rest of the paper is organized as follows. Section \ref{sec:GPE} reviews the principal ideas of the GP regression models proposed in the literature. Section \ref{sec:meth}  develops our GP-Huber regression model and demonstrates its robustness using a numerical example. Section \ref{sec:examples} evaluates its  performance on a  numerical example and real-world applications. Section 5 concludes the paper and discusses future work. This paper has supplementary materials that are available online.

\section{BACKGROUND}
\label{sec:GPE}
This section reviews the general structure of the Gaussian process models for regression proposed in the literature.  

In the Bayesian formulation, the prior distribution over the parameter vector, $\mathbf{\eta}$, of the probabilistic model ${\mathcal{H}}$ captures our prior knowledge of the process thought to be a function of the $d$ - dimensional regressor vector, $\mathbf{x}$. As for the posterior distribution, it is conditioned on both the prior distribution and  $n$ observations of the unidimensional output variable contained in the vector, $\mathbf{y}$, which models the residual epistemic uncertainty, primarily due to the lack of observations and limited understanding of the underlying physical process. Formally, the posterior probability  is  given by
\begin{equation}\label{000001}
   P(\bm{\eta}|\mathbf{y},\mathbf{X},\mathcal{H})=\frac{P(\mathbf{y}|\mathbf{X},\bm{\eta},\mathcal{H})P(\bm{\eta}|\mathcal{H})}{P(\mathcal{D}|\bm{\eta},\mathcal{H})}, 
\end{equation}
where $\mathbf{X}\in\mathbb{R}^{n\times d}$ is a  matrix consisting of $n$ observations of the input vector, $\mathbf{x}$, and $\mathcal{D}$ denotes the training data, $(\mathbf{y},\mathbf{X})$. In Gaussian process models, the systematic dependency between the input and output vectors is given by a latent function, $f(\mathbf{x}):\mathbb{R}^{d}\rightarrow \mathbb{R}$. In a truly non-parametric sense, the latent vector function, $\mathbf{f}=[{f}(\mathbf{x}_{1}),\hdots,{f}(\mathbf{x}_{n})]^{T}$, do not take any parametric form, but instead, it is assumed to have a prior probability distribution while following a joint multivariate normal distribution  with mean vector, $\mathbf{m}(\mathbf{X})=[m(\mathbf{x}_{1}),\hdots,m(\mathbf{x}_{n}))]^{T}$ and covariance matrix, $\mathbf{K}(\mathbf{X},\mathbf{X})$, that is,  
\begin{equation}
    \mathbf{f}|\mathbf{X},\bm{\theta}\sim \mathcal{N}(\mathbf{f}|{\mathbf{m}(\mathbf{X}),\mathbf{K}(\mathbf{X},\mathbf{X})}).
\end{equation}
The covariance matrix, $\mathbf{K}(\mathbf{X},\mathbf{X})$, is a positive semi-definite matrix that captures residual spatial association. The elements of this matrix are kernel functions given by ${K}_{i,j}=k(\mathbf{x}_{i},\mathbf{x}_{j})$, $i,j=1,2,\hdots,n$, where $k(\mathbf{x}_{i},\mathbf{x}_{j})$ is typically chosen from a parametric kernel family such as the Gaussian or the Matern kernel. These parameters are known as hyperparameters of that kernel function denoted by $\bm{\theta}$.
The mean function structure, $m(\mathbf{x})$, can be elicited with a basis vector function, $\bm{h}(\mathbf{x}):\mathbb{R}^{d}\rightarrow\mathbb{R}^{p}$, as $m(\mathbf{x})=\bm{h}(\mathbf{x})^{T}\bm{\beta}$ when the form of $\mathbf{f}$ is known. It is assumed to be zero for further computational simplification. Therefore, assuming that the prior probability distribution of $\mathbf{f}$ is Gaussian with zero mean and covariance matrix, $\mathbf{K}$, yielding the probability density function, $p_{G}(\mathbf{f}|\mathbf{0},\mathbf{K})$, the posterior probability density function of $\mathbf{f}$ using the Baye's rule in \eqref{000001} is given by 
\begin{equation}
    p(\mathbf{f}|\mathcal{D,\bm{\phi},\bm{\theta}})=
    \frac{p(\mathbf{y}|\mathbf{f},\bm{\phi})p_{G}(\mathbf{f}|\mathbf{0},\mathbf{K})}{p(\mathcal{D}|\bm\phi,\bm{\theta})},
\end{equation}
where the likelihood of the observation vector, $\mathbf{y}$, is expressed by 
\begin{equation}
    p(\mathbf{y}|\mathbf{f},\bm{\phi})=\prod_{i=1}^{n}p(y_{i}|f(\mathbf{x}_{i}),\bm{\phi}).
\end{equation}
The evidence, also known as the marginal likelihood, is denoted as $p(\mathcal{D}|\bm{\phi},\bm{\theta})$.

In the simplest case of the GP regression model, commonly known as the conjugate model, the error, $\epsilon_{i}$, in an output observation, $y_
{i}$, is assumed to be Gaussian and additive, that is, $y_
{i} = {f}(\mathbf{x}_{i}) + \epsilon_{i}$, where $\epsilon_{i} \sim \mathcal{N}(0,\sigma^{2}(\mathbf{x}_{i}))$ is a homoscedastic independent and identically distributed (i.i.d.)  random variable with constant variance, $\sigma^{2}(\mathbf{x}_{i}) = \sigma^{2}$, which is the likelihood hyperparameter, $\phi=\sigma^{2}$. 
Given a form for $k(\cdot,\cdot)$, the observation vector, $\mathbf{y}$, follows a multivariate Gaussian distribution expressed as
\begin{equation}\label{eq4}
    \mathbf{y}|\mathbf{f},\mathbf{X},\sigma^{2},\bm{\theta}\sim\mathcal{N}\left(\mathbf{y}|\mathbf{f},\mathbf{K}{(\mathbf{X},\mathbf{X}}|\bm{\theta})+\sigma^{2}\mathbf{I}_{n}\right).
\end{equation}
For instance, for a Gaussian kernel, ${k}(\mathbf{x}_{i},\mathbf{x}_j)=\tau^{2}\textrm{exp}\,\left(-\sum_{k=1}^{d}\frac{(\textrm{x}_{ik}-\textrm{x}_{jk})^2}{{\bm{s}}_k^2}\right)$ and $\bm{\theta}=[\tau,{s}_{1},{s}_{2},\hdots,{s}_{n}]^{T}$ is the the hyperparameter vector of kernel function $k(\cdot,\cdot)$. 

Using the multivariate Gaussian conditional identities, we obtain  key predictive mean vector and covariance matrix estimate at $n^{*}$ new test points contained in $\mathbf{X}^{*}$, which are respectively  expressed as
\begin{equation}\label{eq10}
 \bm{\mu}^{*}(\mathbf{X}^{*}) = {\mathbf{C}}(\mathbf{X}^{*})^{T}{\mathbf{R}}^{-1}\mathbf{r};
\end{equation} 
\begin{equation}\label{eq11}
    {\bm{\Sigma}}^{*}(\mathbf{X^{*}})={\mathbf{V}}(\mathbf{X}^{*})-{\mathbf{C}}(\mathbf{X}^{*})^{T}{\mathbf{R}}^{-1}{\mathbf{C}}(\mathbf{X}^{*}),
\end{equation}
where $\mathbf{r}=\mathbf{y}-\mathbf{f}$ is the residual vector and ${\mathbf{R}}=\mathbf{K}(\mathbf{X},\mathbf{X}|{\bm{\theta}})+{\sigma}^{2}\mathbf{I}_{n}$.  Here,  ${\mathbf{C}}(\mathbf{X}^{*})=\bm{k}(\mathbf{X},\mathbf{X}^{*}|{\bm{\theta}})$ is the covariance matrix of the training and the testing points and ${\mathbf{V}}(\mathbf{X}^{*})=\bm{k}(\mathbf{X}^{*},\mathbf{X}^{*}|{\bm{\theta}})+{\sigma}^{2}\mathbf{I}_{n^{{*}}}$ is the covariance matrix of the testing points. The hyperparameters of the likelihood function $\sigma^{2}$ and kernel function $\bm{\theta}$ are estimated in an ML-II type framework in which the evidence is maximized. Formally, we have 
\begin{equation}
    (\widehat{\sigma}^{2},\widehat{\bm{\theta}})=\underset{{\sigma}^{2},{\bm{\theta}}}{\mathrm{arg\, max}}\; p(\mathcal{D}|\sigma^{2},\bm{\theta}).
\end{equation}
The above model is known as the conjugate model since the likelihood function is normal along with the normal prior on the latent vector function, $\mathbf{f}$, which yields predictive posterior distribution with analytically tractable mean and covariance functions. In the cases of non-normal noise distributions, the posterior distributions of  $\mathbf{f}$ carry heavier tails than the normal distribution. Therefore, the Bayesian inference with Gaussian likelihood is not robust against outliers occurring in real data. Therefore, to make robust predictions in the Gaussian process framework, the Gaussian likelihood explaining the observations is replaced by either the Student's t-distribution   (\cite{Vanhatalo2009}, \cite{Hartmann2019}, \cite{Kuss2005}), or the Laplace distribution (\cite{Kuss2006}), or a heavy-tailed Gaussian mixture distribution  (\cite{Kuss2006}). With these likelihood functions, Bayesian inference becomes complicated and results in intractable analytical forms of the posterior distribution of $\mathbf{f}$. Consequently, various inference techniques have been proposed in the literature, which are commonly classified into two types: (i) posterior probability distribution of $\mathbf{f}$ that is approximated by a normal distribution with individual variances and (ii) MCMC methods that directly sample from the posterior distribution. Special care needs to be taken for the latter to make the samples uncorrelated for better mixing the chains.

\section{Gaussian Process with Huber Density Function\label{sec:meth}}
In this section, we derive the GP-Huber regression model and further develop Bayesian approximation methods to estimate the latent vector function, $\mathbf{f}$, and the hyperparameters of the likelihood and covariance function, $(\bm{\phi},\bm{\theta})$.

\subsection{GP-Huber Regression Model} 
\cite{Huber1964} proposed a loss function that is a truncated mixture of two commonly used loss functions: squared loss, $\mathcal{L}(r)=r^{2}$, and absolute loss, $\mathcal{L}(r)=|r|$, which is commonly known as Huber loss in robust statistics. The associated least favorable Huber density function with a fraction of contamination $\varepsilon$ is defined as 
\begin{equation}\label{003}
    p_{H}(\mathbf{y}|\mathbf{f},{\phi})= \prod_{i=1}^{n}\frac{1-\varepsilon}{\sqrt{2\pi}w_{i}\sigma s}\textrm{exp}\left(-\rho(r_{S_i})\right),
\end{equation}
where  $\rho(\cdot)$ denotes the Huber loss function given by
\begin{equation}
    \rho(r_{S_i})=\begin{cases}
        \frac{r_{S_i}^{2}}{2} & |r_{S_i}|\leq b,\\
       b(|r_{S_i}|-\frac{b}{2}) & |r_{S_{i}}|>b.
    \end{cases}
\end{equation}
This function is strongly convex in a uniform neighborhood of its minimum at $r=0$. The threshold, $b$, is typically set to $1.5$ to achieve high efficiency at the Gaussian distribution. The standardized residuals, $r_{S_i}=\frac{y_{i}-f(\mathbf{x}_{i})}{w_{i}\sigma s}$, are scaled by the weights, $w_{i}$, based on projection statistics.

The projection statistics are defined as the maxima of the standardized projection distances obtained by projecting the point cloud in the directions that originate from the coordinate-wise median and that pass through each of the data points, $\mathbf{x}_{i}$ (see, \cite{Mili1996}). Formally, we have 
 \begin{equation}\label{eq36}
     {\textrm{PS}}(\mathbf{x}_{i})=\underset{||\bm{u}_{j}||=1}{max}\; \frac{|\mathbf{x}_i^{T}\bm{u}_{j}-\underset{k}{\textrm{median}}(\mathbf{x}_k^{T}\bm{u}_{j})|}{1.4826\;\underset{i}{\textrm{median}}\;|\mathbf{x}_{i}^{T}\bm{u}_{j}-\underset{k}{\textrm{median}}(\mathbf{x}_{k}^{T}\bm{u}_{j})|},
 \end{equation}
where $\bm{v}_j=\mathbf{x}_{j}-\mathbf{M}$ and  $\bm{u}_j=\frac{\bm{v}_{j}}{||\bm{v}_{j}||}$ for  $j,k=1,\hdots,n$. Here, $\mathbf{M}$ denotes the coordinatewise median given by 
 \begin{equation*}
     \mathbf{M}=\{\underset{j=1,\hdots,n}{\textrm{med}}\;\mathbf{x}_{j1},\hdots,\underset{j=1,\hdots,n}{\textrm{med}}\;\mathbf{x}_{jd}\}.
 \end{equation*}
The projection statistics are fast to calculate; but this comes at the expense of a loss of affine equivariance, which is not a shortcoming for our method because they are used as a robust diagnostic tool to identify outliers. 
We define the weights used in the  standardized residual $r_{S_i}$ as
\begin{equation}\label{15}
  {w}(\mathbf{x}_i)=\begin{cases}
    1,& \textrm{for\;} {\textrm{PS}^{2}_i}\leq c_{i},\\
    \frac{c_{i}}{\textrm{PS}_{i}^{2}}, & \textrm{for\;}{\textrm{PS}_i^{2}}>c_{i},
\end{cases}
\end{equation}
which downweight outliers. \cite{Rousseeuw1991} and \cite{Mili1996} showed that, when $n>5d$, the squared projection statistics, $\textrm{PS}^{2}(\mathbf{x}_{i})$, roughly follow a $\chi^{2}$ distribution with a degree of freedom equal to the number of non-zero elements $\nu_{i}$ in the row vector of the associated regressor, $\mathbf{x}_{i}$, i.e., $\textrm{PS}^{2}(\mathbf{x}_{i})\sim \chi^{2}_{\nu_{i}}$. However, when $n\leq5d$, it is the projection statistics that roughly follow a $\chi^{2}$ distribution, that is,  
$\textrm{PS}(\mathbf{x}_{i})\sim \chi^{2}_{\nu_{i}}$. Consequently, the threshold $c_{i}$ is chosen as the $97.5$ percentile of the chi-square distribution with $\nu_{i}$ degrees of freedom  for an individual input point, $\mathbf{x}_{i}$. The robust estimator of the scale of the residuals, $s=1.4826 b_{d}\; \textrm{median}|\bm{r}|$, accounts for the unknown error variance (see, \cite{Mili1996}) and $b_{d}=1+{5}/{(n-d)}$ is the dimensionality correction factor. \cite{Yohai1995} showed that the projection statistics attain the maximum breakdown point given by ${[(n-d-1)/2]}/{n}$. Note that we set the fraction of contamination to $0.45$, $\varepsilon=0.45$, and the scale $s$ is calculated in every iteration of the inference. Therefore, the only hyperparameter of the likelihood function that needs to be inferred is ${\phi}=\sigma^{2}$.

\begin{figure}[t]%
\centering
\subfloat[\centering  ]{{\includegraphics[height=5cm,width=5.7cm]{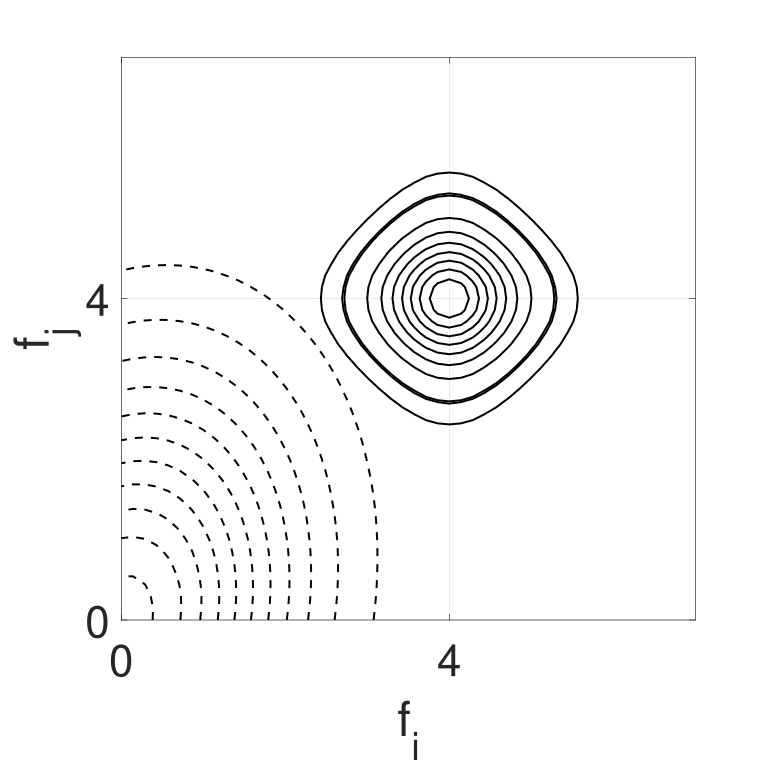}}}%
\quad \quad \quad 
\subfloat[\centering ]{{\includegraphics[height=5cm,width=5.7cm]{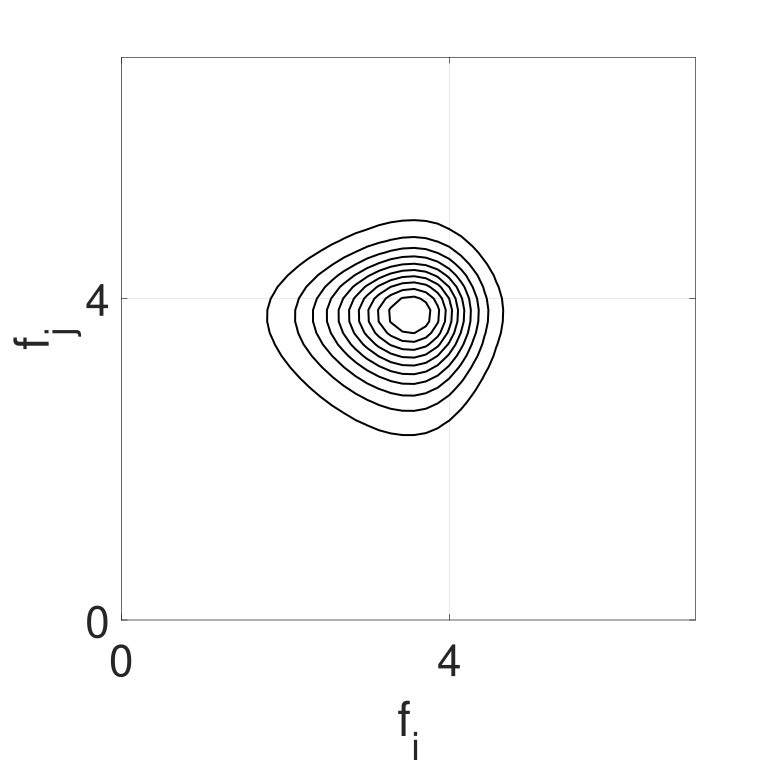} }}%
\caption{\textit{Plots of the contours of the Huber likelihood function showing its convexity: (a) Prior Huber distribution, (b) Unnormalized posterior Huber distribution.}}%
    \label{Ill}%
\end{figure}

Assuming that the latent vector function, $\mathbf{f}$, follows the normal prior probability distribution, that is,
\begin{equation}
\mathbf{f}|\mathbf{X},\bm{\theta}\sim \mathcal{N}(\mathbf{f}|\mathbf{0},\mathbf{K}),
\end{equation}
then, its posterior probability density function will be given by
\begin{equation}
    p(\mathbf{f}|\mathcal{D},\bm{\theta},{\sigma})=\frac{p_{G}(\mathbf{f}|\mathbf{0},\mathbf{K})}{p(\mathcal{D}|\bm{\theta},{\sigma})} p_{H}(\mathbf{y}|\mathbf{f},{\sigma}),
\end{equation}
where p$_{G}(\mathbf{f}|\mathbf{0},\mathbf{K})$ is the normal probability density function of $\mathcal{N}(\mathbf{f}|\mathbf{0},\mathbf{K})$.  As for the evidence, it is defined as  
\begin{equation}\label{eq040}
   p(\mathcal{D}|{\sigma},\bm{\theta})=\int p_{G}(\mathbf{f}|\mathbf{0},\mathbf{K}) p_{H}(\mathbf{y}|\mathbf{f},{\sigma})d\mathbf{f}.     
\end{equation}
The resulting posterior distribution of $\mathbf{f}$ is unimodal since, as illustrated in Figure \ref{Ill}, the Huber density function is log-concave  around $\mathbf{y}=[4 ;4 ]$, where the prior distribution of $\mathbf{f}$ is sampled from the normal distribution, $\mathcal{N}\left(\begin{bmatrix}
    0 \\0
\end{bmatrix},\begin{bmatrix} 2 & 1\\ 1& 4   \end{bmatrix}\right)$.  The marginal likelihood or evidence, $p(\mathcal{D}|\sigma,\bm{\theta})$, and the posterior distribution of  $\mathbf{f}$, $p(\mathbf{f}|\mathcal{D}, \sigma, \bm{\theta})$, are not anymore analytically tractable and they require the application of an approximate method of inference.

In other robust GP regression models such as those based on Student's t-likelihoods proposed by \cite{Kuss2006}, a scale-mixture representation of the Student's t-distribution is utilized and a variational approximation is implemented using the presumption that the likelihood is Gaussian and characterized by the individual variances representing the scale-mixture. Combined with the Kullback-Leibler divergence, $\textrm{KL}(q||p)$, between the true posterior, $p$, and the approximation, $q$, an expectation maximization (EM)-type algorithm is implemented. As for the models with Laplace likelihoods, a scale mixture of normal distributions is utilized. Since the resulting posterior is a unimodal distribution, the EP approximation and the MCMC sampling can be implemented. Note that, here, a Laplace approximation is inappropriate because the discontinuous derivatives of the Laplace likelihood at zero may cause the Hessian matrix to be undefined.

\subsection{Approximate Bayesian Inference}
Regarding the Bayesian inference of the proposed GP with the Huber likelihood function, we develop the Laplace approximation method and hybrid Monte Carlo sampling. The scale-mixture representation of the Laplace distribution is utilized for the former.     
\subsubsection{Laplace Approximation}
To ensure the continuity of the derivative with respect to the latent vector function,  $\mathbf{f}$, we utilize the pseudo-Huber loss function (see, \cite{Charbonnier1997}) defined as 
\begin{equation}
    \rho(r_{S})=b^{2}\left(\sqrt{(1+\left(\frac{r_{S}}{b}\right)^{2}} -1\right).
\end{equation} 
This function approximates $r_{S}^{2}/2$ for small values of the standardized residual, $r_{S}$, and a straight line with slope $b$ for large values of $r_{S}$. Laplace approximation of the posterior requires the likelihood to be log-concave in order for it to be represented by a unimodal multivariate normal distribution. It is executed by approximating the posterior distribution of  $\mathbf{f}$ with a normal distribution  (see, \cite{Rue2009}), that is, 
\begin{equation}
    \mathbf{f}|\mathcal{D},\sigma,\bm{\theta} \sim \mathcal{N}(\hat{\mathbf{f}}|\mathbf{f},\mathbf{A}).
\end{equation}
A Taylor series expansion about the largest mode of the un-normalized posterior density function of $\mathbf{f}$ yields $q(\mathbf{f}|\mathcal{D},\sigma, \bm{\theta})\approx p_{H}(\mathbf{y}|\mathbf{f},\sigma)p_{G}(\mathbf{f}|\mathbf{0},\mathbf{K})$. The latter is used to define the MAP estimate, $\hat{\mathbf{f}}$, of $\mathbf{f}$,  given by 
\begin{equation}
    \hat{\mathbf{f}}=\underset{{{\mathbf{f}}}}{\mathrm{arg\, max}}\; \textrm{ln}\;q(\mathbf{f}|\mathcal{D},\sigma, \bm{\theta}),
\end{equation}
which may converge to a local mode in case of multimodal likelihood. As for the posterior covariance matrix, $\mathbf{A}$, it is given by 
\begin{equation}
    \mathbf{A}=(\mathbf{K}^{-1}+\mathbf{W})^{-1},
\end{equation}
where $\mathbf{W}$ is equal to the negative of the Hessian of the log-likelihood given by the natural log of \eqref{003}, that is,
\begin{equation}
    \mathbf{W}=-\nabla \nabla_{{\mathbf{f}}} \textrm{ln}\left(p_{H}(\mathbf{y}|\hat{\mathbf{f}},{\sigma})\right).
\end{equation}
Regarding the hyperparameter vector, $(\sigma,\bm{\theta})$, it is estimated by maximizing the log of the approximate evidence given by \eqref{eq040} using the gradient descent or the conjugate gradient method since the gradient can be analytically derived. Formally, we have 
\begin{equation}
    (\hat{\sigma},\hat{\bm{\theta}})= \underset{(\sigma,\bm{\theta})}{\mathrm{arg\, max}}\; \textrm{ln}\; q(\mathcal{D}|\sigma,\bm{\theta}),
\end{equation}
where $q(\mathcal{D}|\sigma,\bm{\theta})\approx p(\mathcal{D}|\sigma,\bm{\theta})$ is the approximate log evidence given by
\begin{equation}
    \textrm{ln}\;q(\mathcal{D}|\sigma,\bm{\theta})=\textrm{ln}
     \;p_{H}(\hat{\mathbf{f}}|\mathbf{f})-\frac{1}{2}\textrm{ln}|\mathbf{K}|-\frac{1}{2}\mathbf{f}^{T}\mathbf{K}^{-1}\mathbf{f}+\frac{1}{2}\textrm{ln}|\mathbf{A}|.
\end{equation}
For more details on Laplace approximation, the reader is referred to \cite{Kuss2005}.

\subsubsection{Hybrid Monte Carlo}
The Huber density function is a mixture of a truncated normal  and a Laplace density function for an absolute standardized residual respectively lying within and outside the threshold $b$. This yields 
\begin{equation}
    \rho(r_{S_i})=\begin{cases}
        \frac{C_{1}}{\sqrt{2\pi}w_{i}\sigma_{g} s}\textrm{exp}\left(-\frac{r_{i}^{2}}{2 w_{i}^{2}\sigma_{g}^{2} s^{2}}\right) & |r_{S_i}|\leq b,\\
        \frac{C_2}{2 w_{i} a s}\textrm{exp}\left(-\frac{b|r_{i}|}{w_{i} a s}\right) & |r_{S_i}|>b,
    \end{cases}
\end{equation}
where $C_{1}$ and $C_{2}$ are the constants respectively defined as $C_{1}=1-\varepsilon$ and $C_{2}=\sqrt{\frac{\pi}{2}}\textrm{exp}(b^{2}/2)$ . The Laplace distribution of $y_{i}$, $ \textrm{Laplace}(y_{i}|f(\mathbf{x}_{i}), a),$ can be represented as a scale mixture of normal distributions, $\mathcal{N}(y_{i}|f(\mathbf{x}_{i}),\sigma^{2}_{i})$ where $\sigma^{2}_{i}$ follows an exponential distribution and  $i=1,\hdots,n_{l}$ are the indices of the points associated with the standardized residuals larger than the threshold $b$ hereafter referred to as outlying points; see \cite{Andrews1974} for more details. Formally, we have 
\begin{equation}
     p_{L}(y_{i}|f(\mathbf{x}_{i}),a) =\int p_{G}(y_{i}|f(\mathbf{x}_{i}),\sigma_{i}^{2})p_{E}(\sigma_{{i}}^{2}|\beta) d\sigma^{2}_{i},
\end{equation}
where $p_{L}$ and $p_{E}$ respectively denote the Laplace and the exponential probability density function  and 
$a=1/\sqrt{2\beta}$.  Using this property, we represent the individual standard deviations corresponding to $n_{l}$ outlying training points as $\{\sigma_{l_{1}}, \hdots, \sigma_{l_{n_{l}}}\}$, which are elements of the vector, $\bm{\sigma_{l}}$. The variance associated with $n_{g}$ inlying points is denoted as $\sigma^{2}_{g}$.
Conclusively, the Huber probability density function takes the form 
\begin{equation}\label{eq044}
    \mathbf{y}|\mathbf{f},\sigma^{2}_{g},\bm{\sigma}^{2}_{l},{\beta}\sim\begin{cases}
               \prod_{i=1}^{n_{g}}C_{1}\mathcal{N}(y_{i}|f(\mathbf{x}_{i}),\sigma^{2}_{g}) & |r_{S_i}|\leq b ,\\  
               \prod_{i=1}^{n_{l}}C_{2}\mathcal{N}(y_{i}|f(\mathbf{x}_{i}),\sigma^{2}_{l_i}) \textrm{Exponential}(\sigma^{2}_{l_i},\beta) & |r_{S_i}|>b,           
  \end{cases}
\end{equation}
where $n_{g}+n_{l}=n$ is the total number of points in the training dataset. An alternative representation of the likelihood function is given by
\begin{equation}\label{eq046}
    \mathbf{y}_{g},\mathbf{y}_{l}|\mathbf{f}_{g},\mathbf{f}_{l},\sigma^{2}_{g},\bm{\sigma}^{2}_{l}\sim\mathcal{N}\left( \begin{bmatrix}
        \mathbf{y}_{g}|\mathbf{f}_{g}\\
         \mathbf{y}_{l}|\mathbf{f}_{l}\\
    \end{bmatrix}, \begin{bmatrix}
        \mathbf{\Sigma}_{gg}& \mathbf{0}\\
       \mathbf{0} & \mathbf{\Sigma}_{ll}\\
    \end{bmatrix} \right),
\end{equation}
where $\mathbf{\Sigma}_{gg}$ and $\mathbf{\Sigma}_{ll}$ both are diagonal matrices, the former with constant diagonal elements equal to $\sigma^{2}_{g}$ and the latter with diagonal entries $\{{\sigma}^{2}_{{l}_{1}}, \hdots,{\sigma}^{2}_{{l}_{n_{l}}}\}$. Let the hyperparameter vector, $\bm{\sigma^{2}}$, consist of the two vectors, $\bm{\sigma}^{2}_{g}$ i.e. the diagonal entries of the matrix $\bm{\Sigma}_{gg}$ and $\bm{\sigma}^{2}_{l}$. The joint posterior probability density function of $\mathbf{f}$, $\bm{\sigma}^{2}$, and $\bm{\theta}$ is given by
\begin{equation}
    p(\mathbf{f},\bm{\sigma}^{2},\bm{\theta})\propto p(\mathbf{y}|\mathbf{f},\bm{\sigma}^{2})p_{G}(\mathbf{f}|\mathbf{0},\mathbf{K})p(\bm{\sigma}^{2}|\bm{\beta})p(\bm{\beta}|\bm{\zeta})p(\bm{\theta}|\bm{\zeta}),
\end{equation}
We assume that the hyper-hyperparameter vector, $\bm{\beta}$, and the hyperparameter vector, $\bm{\theta}$, with parameters contained in $\bm{\zeta}$ follow the log-uniform distribution. Since the distribution of the variance parameter, $\sigma^{2}_{g}$,  of  $n_{g}$ inlying training points is degenerate, the hyper-hyperparameter vector, $\bm{\beta}=[\beta_{g}, \beta_{l}]^T$, corresponding to the $n_{g}$ points follows a degenerate distribution as well. Therefore, $p(\sigma^{2}_{g}|\beta_{g})$ is a Dirac impulse while $\bm{\sigma}^{2}_{l}|\beta_{l}\sim \textrm{Exponential}(\bm{\sigma}^{2}_{l}|\beta_{l})$.      
The samples generated from this distribution are highly correlated. Therefore, in order to better mix the Monte Carlo chains, we follow \cite{Kuss2006} as follows: 
\begin{equation}
     p(\bm{\sigma}^{2},\bm{\beta},\bm{\theta})\propto\left[\int p_{G}(\mathbf{y}|\mathbf{f},\mathbf{\Sigma}) p_{G}(\mathbf{f}|\mathbf{0},\mathbf{K})d\mathbf{f}\right]p(\bm{\sigma}^{2}|\bm{\beta})p(\bm{\beta}|\bm{\zeta})p(\bm{\theta}|\bm{\zeta}),
\end{equation}
where the covariance matrix of the $n_{g}$ inlying samples and the $n_{l}$ outlying samples is given by $ \mathbf{\Sigma}=\begin{bmatrix}
        \mathbf{\Sigma}_{gg}& \mathbf{0}\\
       \mathbf{0} & \mathbf{\Sigma}_{ll}\\
    \end{bmatrix}$. 
The samples can be used to obtain the approximated probability density functions of the latent vector function, $p(\mathbf{f}_{*}|\mathcal{D},\mathbf{X}_{*})$, at the new test inputs contained in $\mathbf{X}_{*}$ by averaging over all unknowns. Formally, we have 
\begin{equation}
    p(\mathbf{f}_{*}|\mathcal{D},\mathbf{X}_{*})=\int p(\mathbf{f}_{*}|\mathbf{f},\bm{\sigma}^{2},\bm{\theta},\mathbf{X}_{*},\mathcal{D})p(\mathbf{f},\bm{\sigma}^{2},\bm{\theta}|\mathcal{D})d\mathbf{f}d\bm{\sigma}^{2} d\bm{\theta}.
\end{equation}
For $T$ samples, it can be evaluated as
 \begin{equation}
     p(\mathbf{f}_{*}|\mathcal{D},\mathbf{X}_{*},\bm{\zeta})=\frac{1}{T}\sum_{t=1}^{T}\int p(\mathbf{f}_{*}|\mathbf{f},\mathbf{X},\mathbf{X}_{*},\bm{\theta}_{t})p(\mathbf{f}|\mathcal{D},\bm{\sigma}^{2}_{t},\bm{\theta}_{t})d\mathbf{f}. 
 \end{equation}

The trade-off between the breakdown point and the statistical efficiency at the Gaussian distribution of the GP-Huber regression model is addressed by taking advantage of the Huber estimator. This estimator has a bounded influence function and a positive breakdown point, which indicates that it is robust to outliers. Furthermore, it exhibits good statistical efficiency for a broad range of thick-tailed probability distributions of noise.

\section{CASE STUDIES}
\label{sec:examples}
In this section, we first demonstrate the robustness and applicability of the proposed GP regression model with Huber likelihood (GP-Huber) to thick-tailed error distributions on the Neal dataset. We then pose a feature selection problem in order to evaluate its performance on the Friedman dataset. Then, we compare it with the GP regression model with Student's t-likelihood and the GP model with Laplace likelihood using  various approximate Bayesian inference methods such as Laplace approximation, expected propagation, and Markov chain Monte Carlo (MCMC) sampling. The incorporated sampling scheme for the MCMC is a Gibbs sampler for the GP with Student's t-likelihood and a hybrid Monte Carlo sampling for the GP model with Laplace likelihood. We finally apply it to two real-world problems: (i) prediction of  the median value of $1000$ owner-occupied homes in the Boston area and (ii) estimation of the planet-to-star radius ratio of the exoplanet HD 189733 using real recorded spectroscopic data set. 

For clarity, the GP-Huber regression model along with the other considered GP regression models are abbreviated as follows:
\begin{itemize} 
    \item SCtMCMC: GP with a Student's t error model using a  scale mixture representation and solved via the MCMC integration of the latent vector function, $\mathbf{f}$, the hyperparameters of the likelihood function, $\bm{\phi}=(\nu,\sigma^{2})$, and the hyperparameter vector, $\bm{\theta}$. 
    \item SCt4MCMC: GP with Student's t error model with fixed degrees of freedom, $\nu=4$, solved via the MCMC integration over $\mathbf{f}$, $\bm{\phi}=\sigma^{2}$, and $\bm{\theta}$.  
    \item tLA: Student's t error model with the Laplace approximation solved via the Laplace integration over $\mathbf{f}$ and the MAP estimate of the hyperparameters $\bm{\phi}=(\nu,\sigma^{2})$ and $\bm{\theta}$. 
    \item  HuberMCMC: GP with Huber likelihood solved via the  MCMC integration over $\mathbf{f}$, $\bm{\phi}=\sigma^{2}$, and $\bm{\theta}$.  
    \item HuberLA: GP with approximate Huber likelihood solved via the Laplace integration over $\mathbf{f}$ and  the MAP estimates of the hyperparameters $\bm{\phi}=\sigma^{2}$ and $\bm{\theta}$.
    \item GP: Conjugate model, which is GP with normal error where the hyperparameters are $\mathbf{f}$, $\bm{\phi}=\sigma^{2}$, and $\bm{\theta}$.
    \item LaplaceMCMC: GP with Laplace likelihood solved via  the MCMC integration over $\mathbf{f}$, $\bm{\phi}=\sigma$, and $\bm{\theta}$.  
    \item LaplaceEP: GP with Laplace likelihood using EP approximation. 
\end{itemize}
In all the simulations, an anisotropic squared exponential covariance function is used with individual length scale parameters, \{$s_{1},\hdots,s_{d}$\}, for the input dimensions and for which the mean function is assumed to be zero except for the spectroscopy experiment.

\subsection{Case Study 1: Neal Data}
\label{sec:examples1}
\cite{Neal1997} proposed the following artificial model:  
\begin{equation}\label{ep1}
   y= 0.3+0.4x+0.5sin(2.7x) + 1.1/(1+x^2)+e.
\end{equation}  
A sample of $n=100$ points constitutes the training data set, ($\mathbf{X},\mathbf{y}$). The predictions of the vector function, $\mathbf{f}^{*}$, are made at $n^{*}=541$ test input data points contained in $\mathbf{x}^{*}$ over the interval $[-2.7, 5]$. Since the projection statistics require at least a two-dimensional input space, they are calculated on the regressors' vector, $\mathbf{x}$ combined with the column of ones, i.e., on the matrix $\mathbf{H}=[\mathbf{1}, \mathbf{x}]$. Specifically, for a test point $\mathbf{h}_{i}=[1, x_{i}]$, PS$(\mathbf{h}_i)$ is calculated using \eqref{eq36}. The training input point, $x_i$, is flagged as an outlier if the associated weight, $w_{i}=\textrm{min}\left(1,\frac{c}{\textrm{PS}(\mathbf{h}_{i})^{2}}\right)$, has a value less than one. 

\begin{figure}
    \centering
    \includegraphics[height=7cm,width=14cm]{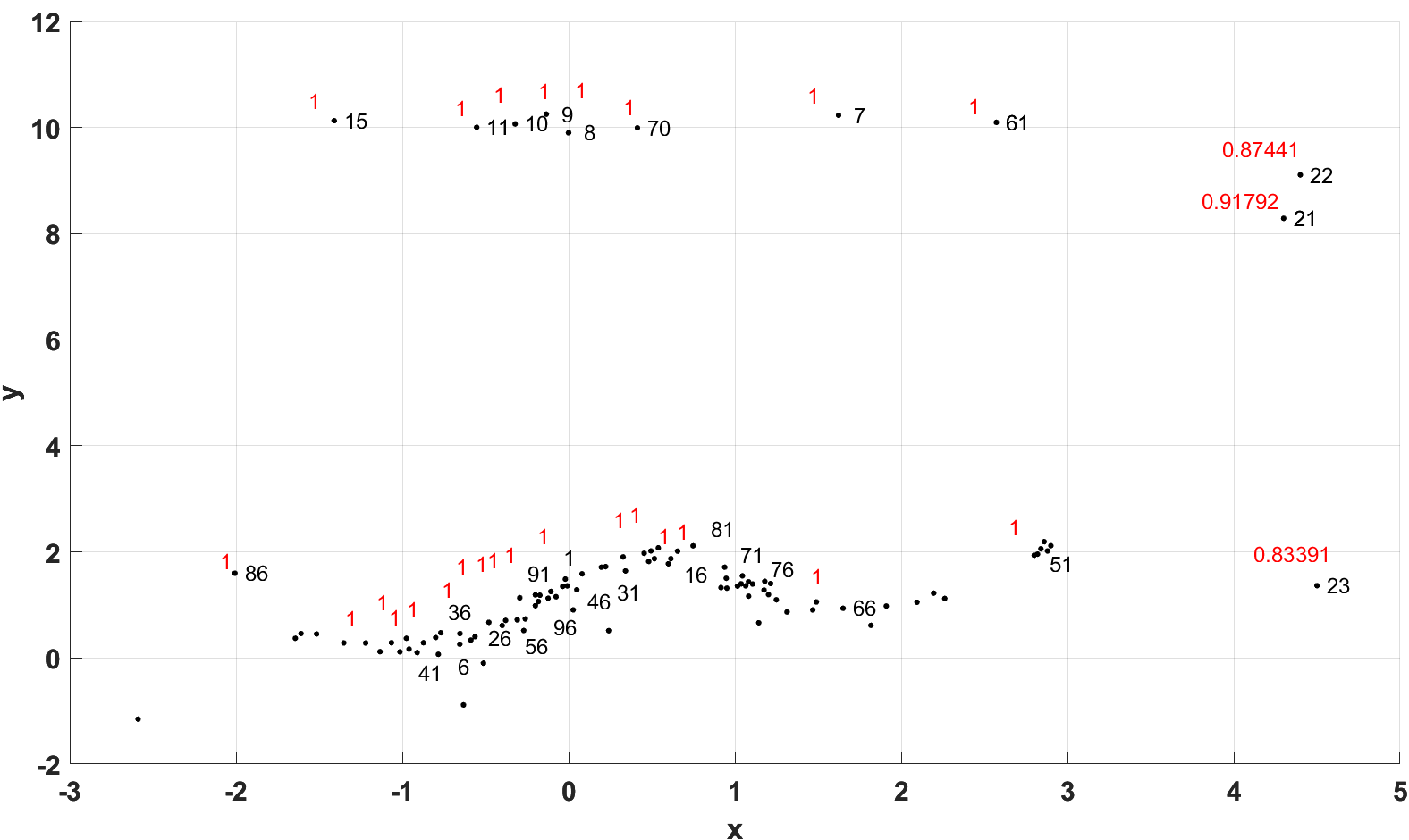}
    \caption{\textit{Weights based on PS for the Neal data. The numbers right to the data points indicate index numbers and the ones to the left in red color indicate the weights associated with that data point.}}
    \label{fig_weights}
\end{figure}

\begin{figure*}%
\centering
\subfloat[\centering ]{{\includegraphics[height=3cm,width=3.6cm]{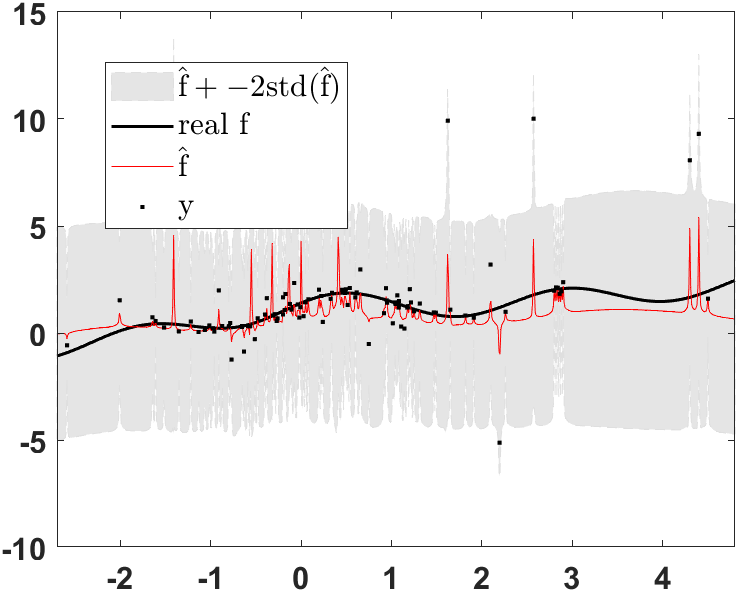}}}%
\subfloat[\centering  ]{{\includegraphics[height=3cm,width=3.6cm]{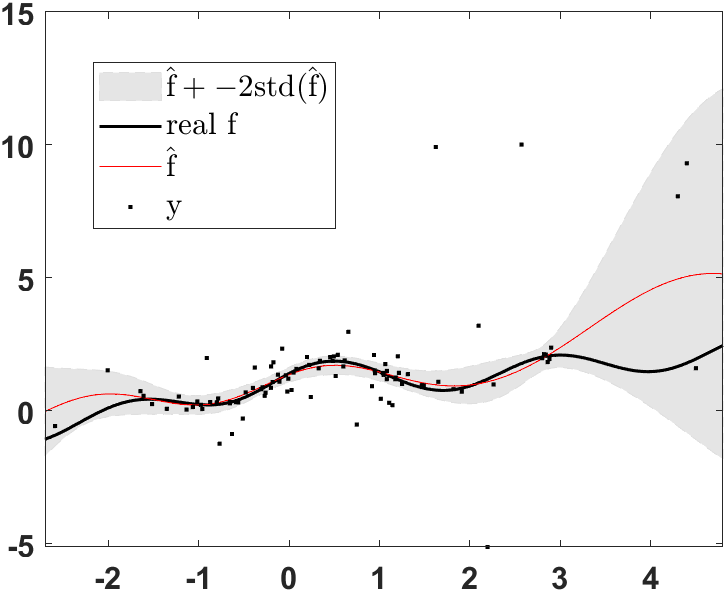}}}%
\subfloat[\centering ]{{\includegraphics[height=3cm,width=3.6cm]{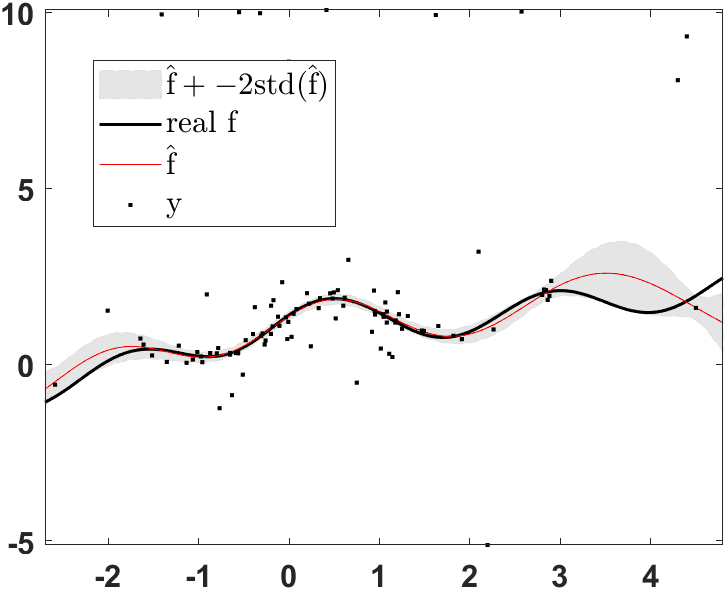}}}%
\subfloat[\centering  ]{{\includegraphics[height=3cm,width=3.6cm]{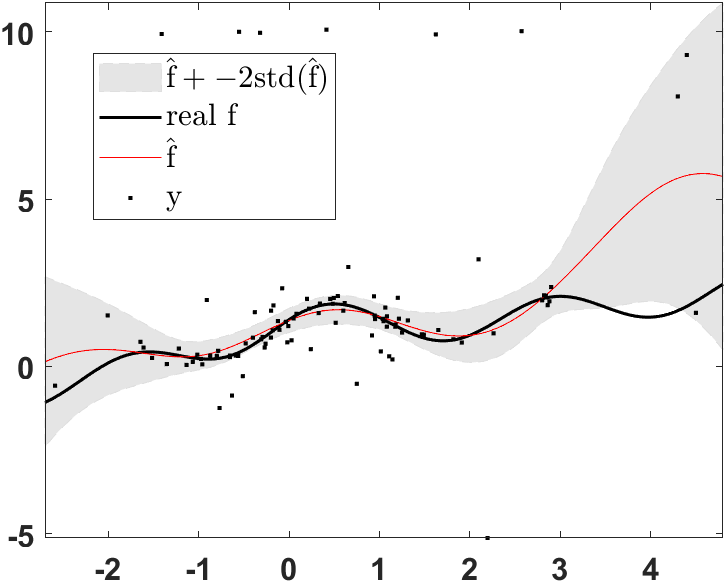}}}%
\;
\subfloat[\centering ]{{\includegraphics[height=3cm,width=3.6cm]{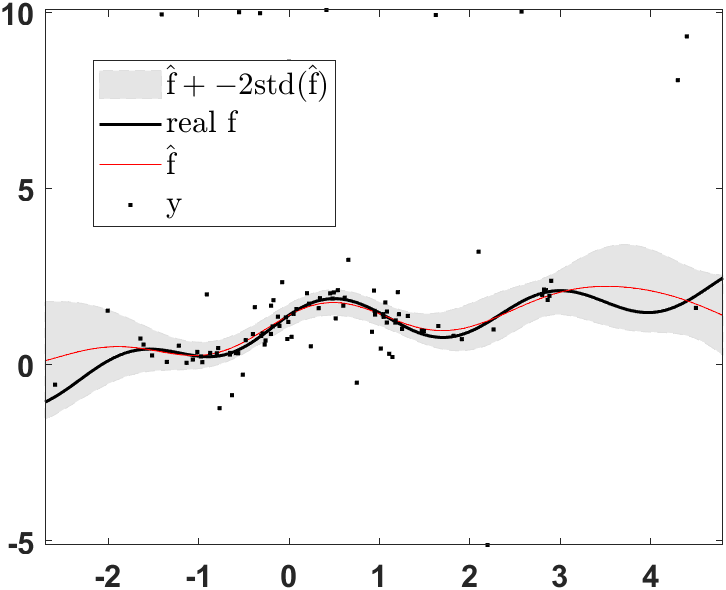} }}%
\subfloat[\centering ]{{\includegraphics[height=3cm,width=3.6cm]{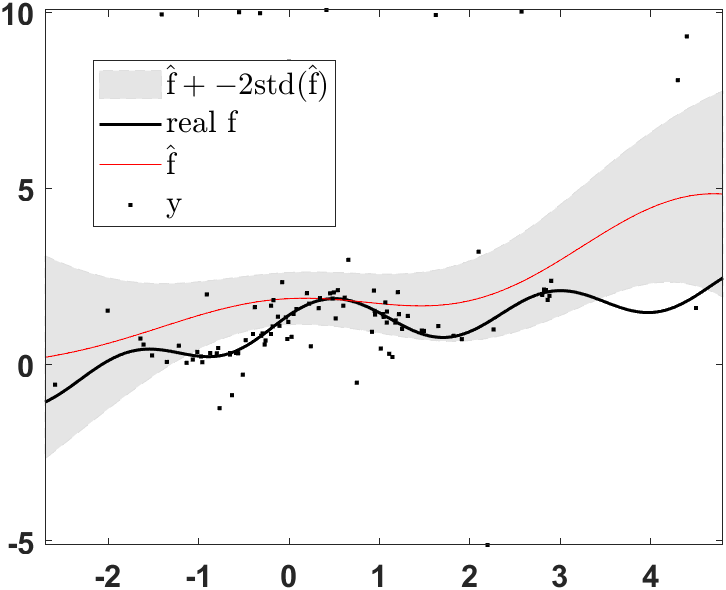} }}%
 \subfloat[\centering ]{{\includegraphics[height=3cm,width=3.6cm]{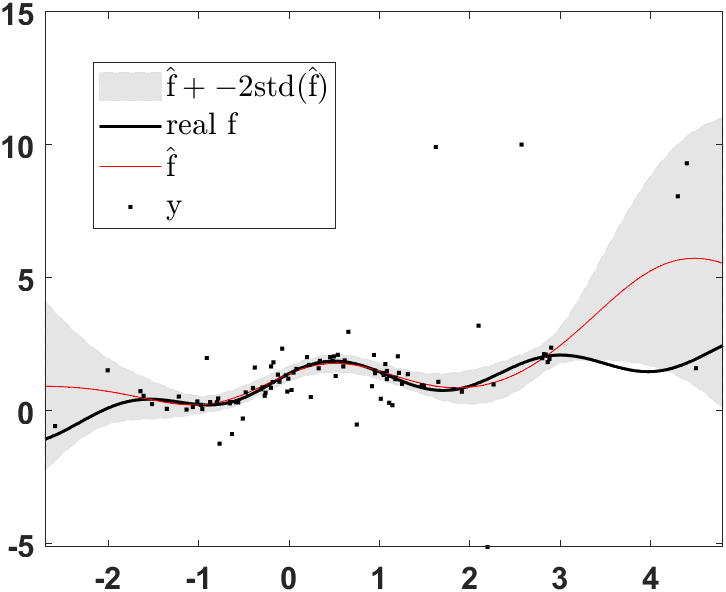}}}%
\subfloat[\centering  ]{{\includegraphics[height=3cm,width=3.6cm]{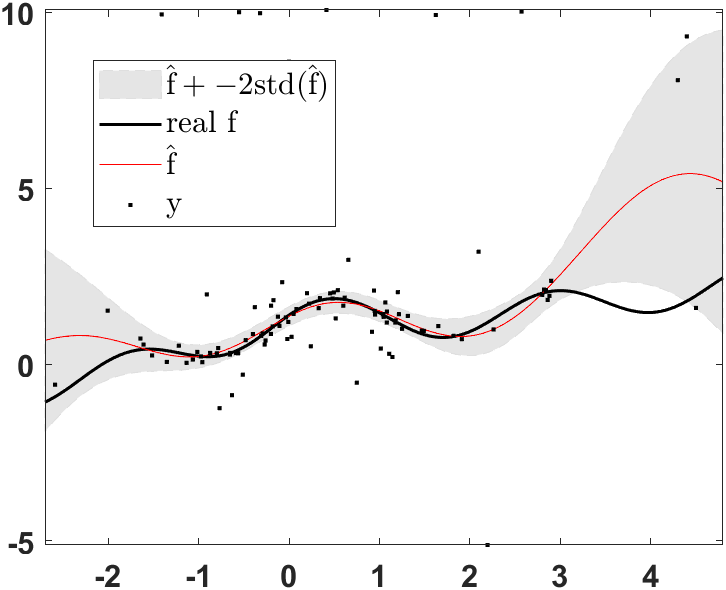} }}%
\caption{\textit{Predicted values for the case of the Student's t-error distribution for the Neal dataset obtained from the eight considered GP regression models: (a) SCtMCMC; (b) SCt4MCMC; (c) tLA; (d) HuberMCMC; (e) HuberLA; (f) GP; (g) LaplaceMCMC; and (h) LaplaceEP.}}%
    \label{Neal_figures}%
\end{figure*}

We demonstrate the proposed GM-Huber in four cases of error probability distribution: (i) $\mathcal{N}(0.01,0.08)$; (ii) the Student's t-distribution with $10$ degrees of freedom;  (iii) $\textrm{Laplace}(0,0.1)$; and (iv) the Cauchy distribution. For each case, we include vertical outliers with a magnitude of 10 in the output training data points, $\mathbf{y}$ indexed $\{7,$ $8,$ $9,$ $10,$ $11,$ $15,$ $61,$ $70\}$. To include bad leverage points, the input points  $\{x_{21},$ $x_{22},$ $x_{23}\}$ are replaced by $\{4.3,$ $4.4,$ $4.5\}$ and the output points $\{y_{21},$ $y_{22},$ $y_{23}\}$ by $\{8.4763,$ $9.1938,$ $0.2833\}$. We also include a group of good leverage points to the input data points $\{x_{50},$ $x_{51},$ $x_{52},$ $x_{53},$ $x_{54},$ $x_{55}\}$, whose corresponding output data point values obtained using \eqref{ep1} are $\{1.9773,$ $2.1271,$ $2.1096,$ $1.8316,$ $1.9467,$ $2.373\}$. We observe that the weights based on the PS corresponding to the bad leverage points are $\{0.9179,$ $0.8744,$ $0.8339\}$ while those corresponding to the good leverage points are equal to $1$ (see Figure \ref{fig_weights}).
The GP-Huber results are accomplished by setting the parameter $b$ of the Huber $\rho$-function to $1.5$ for the Gaussian, the Student's t, and the Cauchy error distributions and to $0.5$ for the Laplace error distribution. We observe that, for the four error probability distributions considered, the GP-Huber with Laplace approximation (HuberLA) model exhibits a lower root mean square error (RMSE), mean absolute error (MAE), and negative log predictive probability (NLP) than the GP-Huber with MCMC integration (HuberMCMC) (see Table \ref{tab_neal}). The models with Student's t-likelihood  with estimated $\nu$ and fixed $\nu$, SCtMCMC and SCt4MCMC, tend to overfit in the Student's t error settings, while the latter exhibits good performance for the Laplace and the Cauchy error distributions (see, Figure \ref{Neal_figures}). We observe in Figure \ref{Neal_figures} that the predicted values obtained from all the considered models except for tLA and HuberLA have an unbounded uncertainty at input test points over the interval $[2.7, 5]$. The performance evaluation scores, namely RMSE, MAE, and NLP are the lowest for tLA over all the parameters. The models with Laplace likelihood, LaplaceMCMC, and LaplaceEP are observed to yield RMSE values not better than the ones obtained using GP. 
\begin{table*}[]
\setlength{\tabcolsep}{2pt}
\centering
\caption{The RMSE, MAE, and NLP for the Neal data.}
  \begin{tabular}{ccccccccc}
  \hline
  {} & \multicolumn{8}{c}{$e\sim \mathcal{N}(0.01,0.08)$}\\
\hline
 Models: & SCtMCMC  &SCt4MCMC &  tLA& HuberMCMC& HuberLA& GP& LaplaceMCMC & LaplaceEP\\
\hline
RMSE & 2.5203&2.4945&0.33005&1.445&0.51252&1.4075&1.4282&1.4319\\
MAE&
1.1145&1.1695&0.1744&0.86049&0.34814&1.1492&0.79683&0.77375\\
NLP&
19.9869&-0.59011&-0.22779&2.4019&-0.286&2.2056&0.42353&0.39064\\
\hline
 {} & \multicolumn{8}{c}{$e\sim \textrm{Student-t}(10)$}\\
\hline
RMSE & 0.75755&1.7524&0.37557&1.4535&0.36876&1.381&1.4846&1.4879\\
MAE &
0.58324&0.93712&0.23536&0.83169&0.24285&1.0864&0.8555&0.86461\\
NLP&
1.776&0.56992&-0.28373&0.69175&-0.031405&1.9524&0.60526&0.72767\\
\hline
 {} & \multicolumn{8}{c}{$e\sim \textrm{Laplace}(0,1)$}\\
\hline
RMSE & 2.5021&0.26366&0.28582&1.1432&0.32524&1.3975&1.4588&1.5091\\ 
MAE&
1.0972&0.14381&0.13801&0.64655&0.18986&1.1395&0.77335&0.80772\\
NLP&
19.3655&-0.7033&-0.5072&0.32566&-0.29686&2.1629&0.28545&0.42087\\
\hline
 {} & \multicolumn{8}{c}{$e\sim \textrm{Student-t}(1)$}\\
\hline
RMSE & 2.4514&0.3396&0.29136&1.325&0.31116&1.4094&1.73&1.4493\\MAE &
1.155&0.19864&0.14681&0.74115&0.1899&1.15&0.96103&0.78332\\
NLP &
15.7405&-0.57424&-0.62215&0.44552&-0.26819&2.2119&0.77819&0.45781\\\hline
\end{tabular}
\label{tab_neal}
\end{table*}

\subsection{Case Study 2: Friedman Data}
\begin{figure}%
\centering
    \subfloat[\centering ]{{\includegraphics[height=5cm,width=7.4cm]{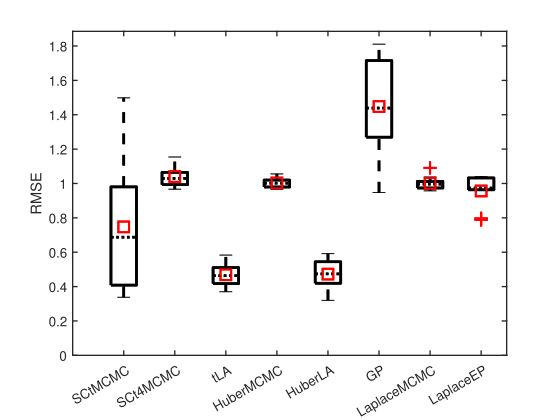}}}%
\subfloat[\centering  ]{{\includegraphics[height=5cm,width=7.4cm]{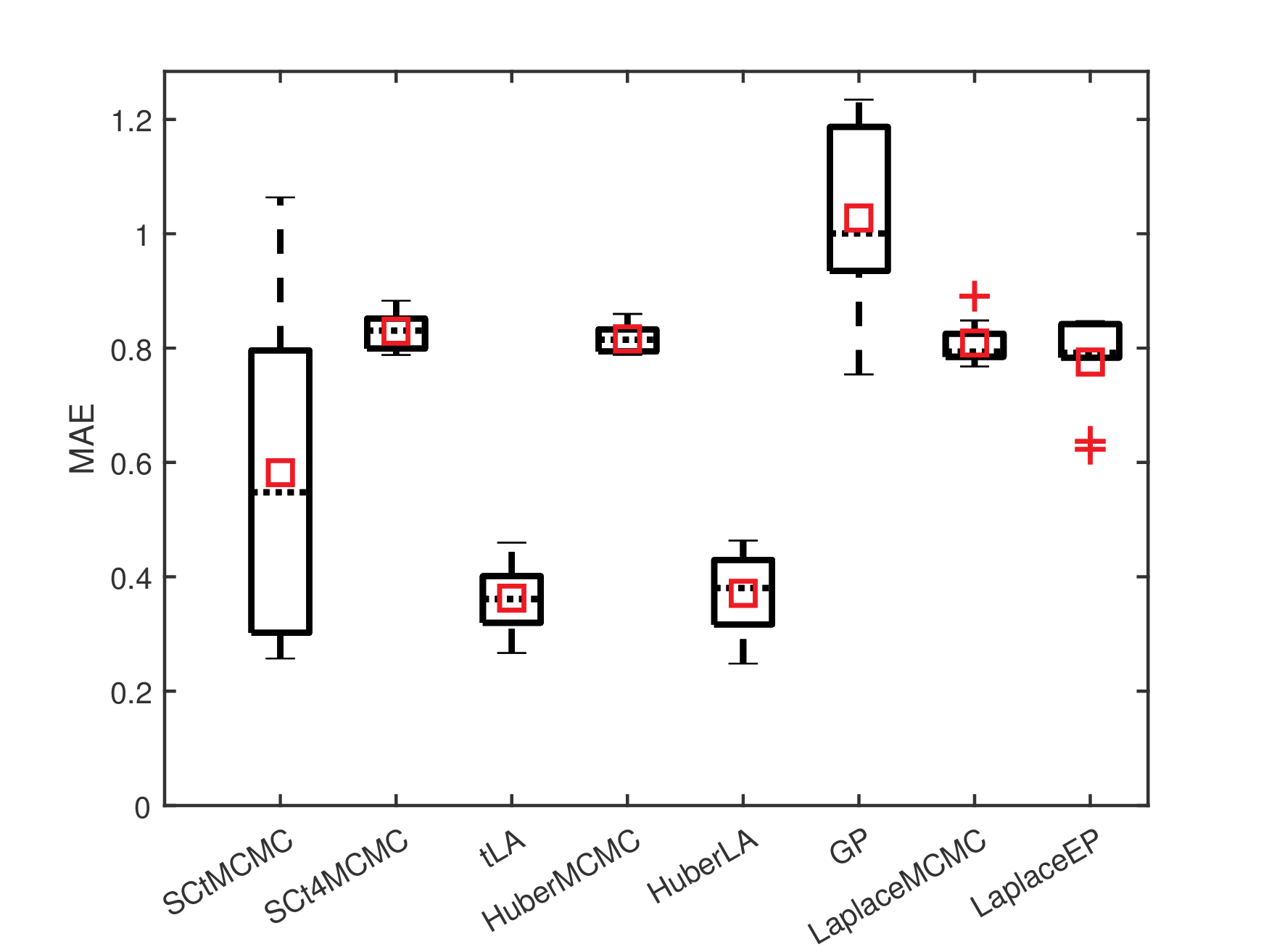}}}%
\;
\subfloat[\centering  ]{{\includegraphics[height=5cm,width=7.4cm]{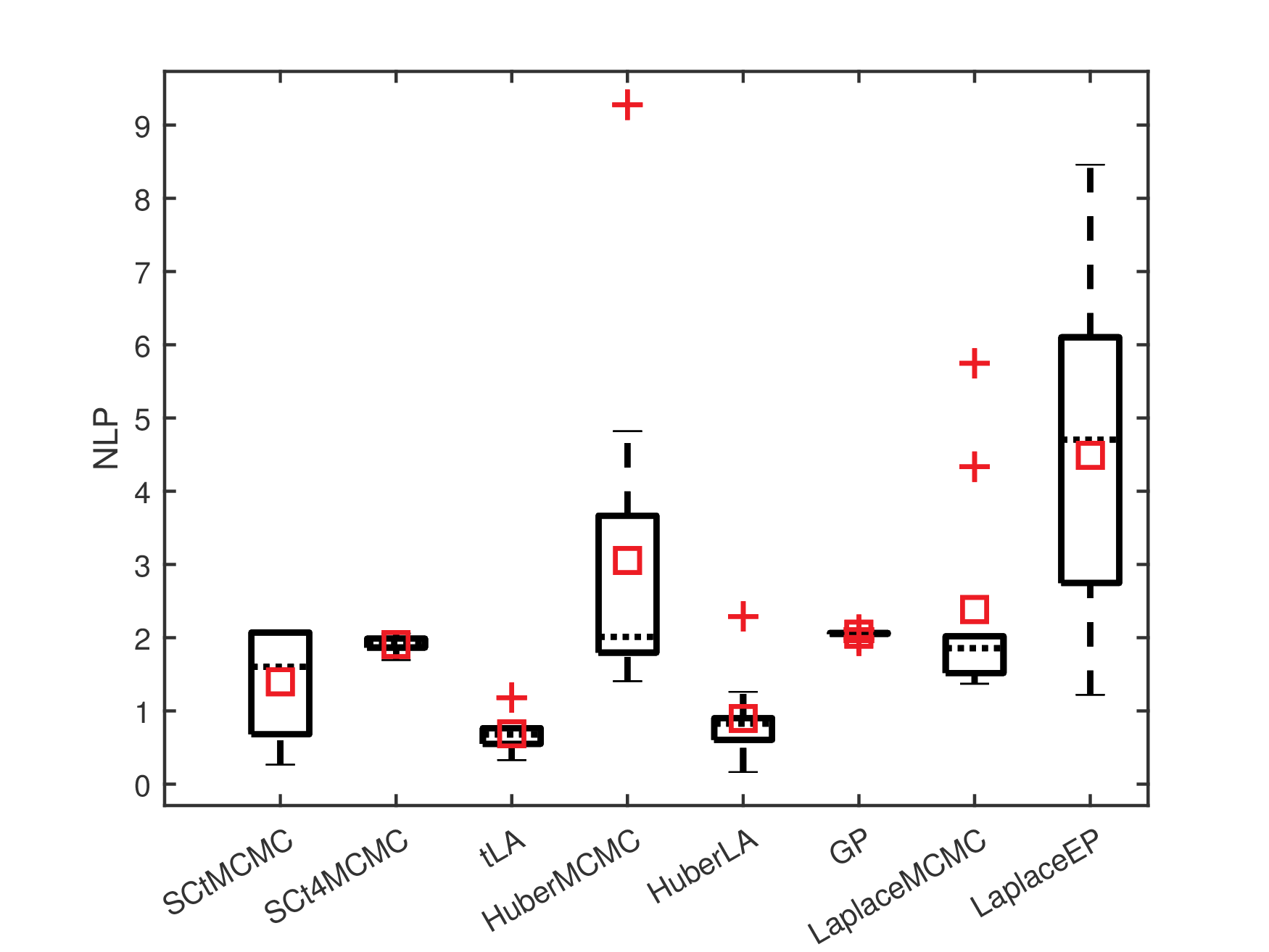}}}%
 \caption{\textit{Box plots of (a) RMSE, (b) MAE, and (c) NLP of the eight considered GP regression models with the normal error distribution, $e\sim \mathcal{N}(0.01,0.08)$, for the Friedman dataset.}}%
 \label{Friedman_studentt}%
\end{figure}

\cite{Friedman1991} proposed a data set that imposes a feature selection problem since the function depends only on the first $5$ input variables of the $10-$dimensional input vector. This dataset is obtained using the following function: 
\begin{equation}
    f(\mathbf{x})=10\textrm{sin}(\pi x_{1} x_{2})+20 ( x_{3}-0.5)^{2}+10x_{4}+5x_{5}.
\end{equation}
Here, we use $10$ training datasets, each with $n=100$, for which the inputs in $\mathbf{x}\in\mathbb{R}^{10}$ are sampled from the uniform distribution $[0,1]^{10}$. The $10$ outliers are randomly included with mean $\mu=10$ and variance $\sigma^{2}_{0}=9$ in all the training datasets. The $n^{*}=10000$ noise-free samples constitute the test dataset. So far, the dataset constructed is similar to the one described in \cite{Kuss2006}. We further include a group of vertical outliers with the magnitude of $10$ at output points, $\{7,$ $8,$ $9,$ $10,$ $11, $15, $61,$ $70\}$, in each training dataset. The fifth dimension of input, $x_{5}$, is arbitrarily chosen to induce bad leverage points at observation points,  $\{21,$ $22,$ $23,$ $24,$ $25,$ $26\}$, with magnitudes $\{8.5312,$ $9.3654,$ $0.7739,$ $0.4802,$ $1.3408,$ $1.7653\}$. The weights based on the PS are calculated as $\{0.1134,$ $0.1214,$ $0.1378,$ $0.1448,$ $0.1255,$ $ 0.1193\}.$ The box plots of the RMSE, MAE, and NLP obtained from the experiments on $10$ training datasets predicted at $n^{*}$ samples are summarized in Figure \ref{Friedman_studentt}. HuberLA is observed to give the lowest RMSE, MAE, and NLP values while their mean (indicated by a square box) and median values are approximately equal to those obtained from tLA. Note that bad leverage points and vertical outliers are included in the training data to assess the robustness of the models under extreme cases that may occur in practice.

\subsection{Case Study 3: Boston Housing Data}
\begin{figure}%
    \centering
    \subfloat[\centering  ]{{\includegraphics[height=8cm,width=15cm]{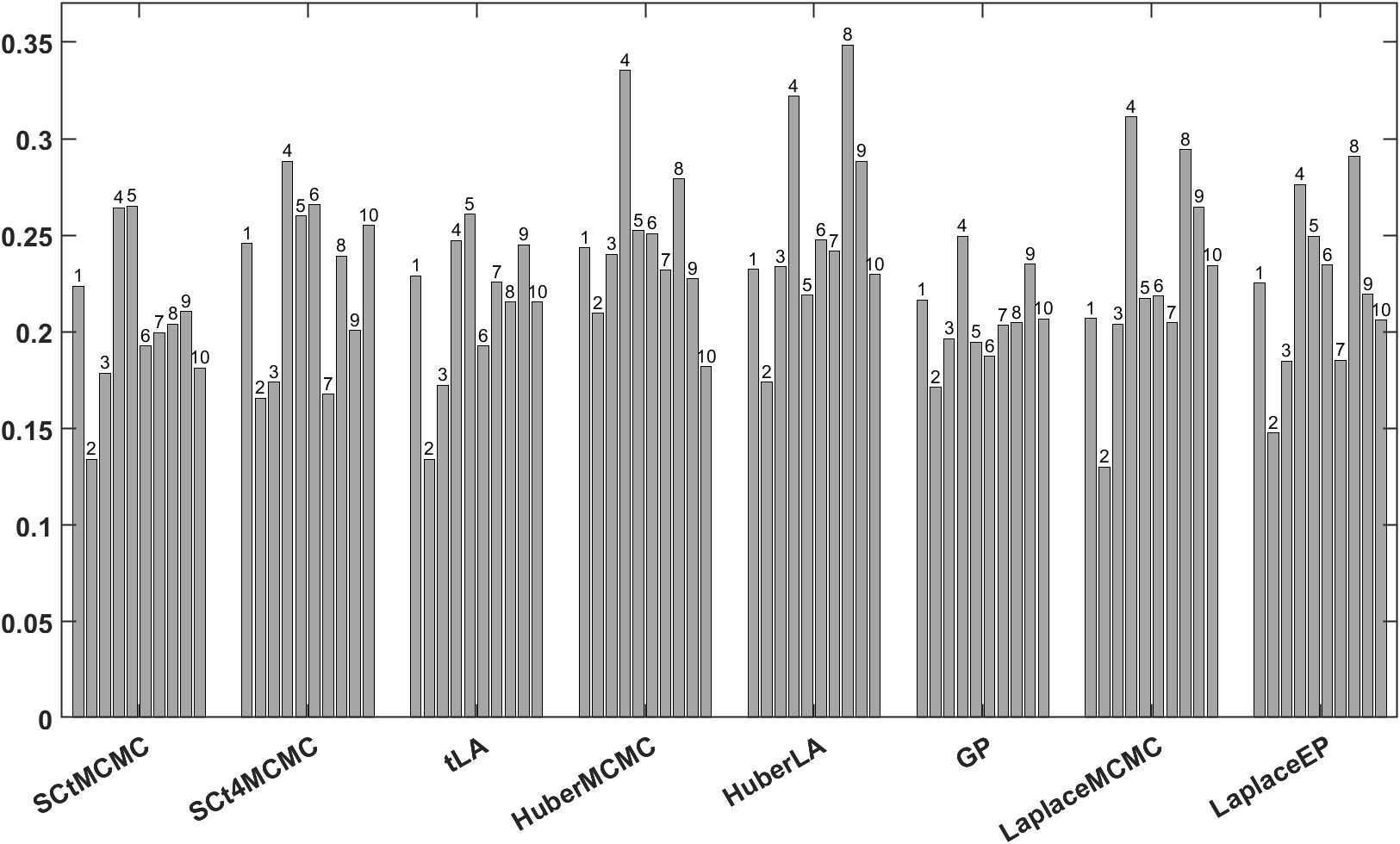} }}%
    \qquad
    \subfloat[\centering ]{{\includegraphics[height=8cm,width=15cm]{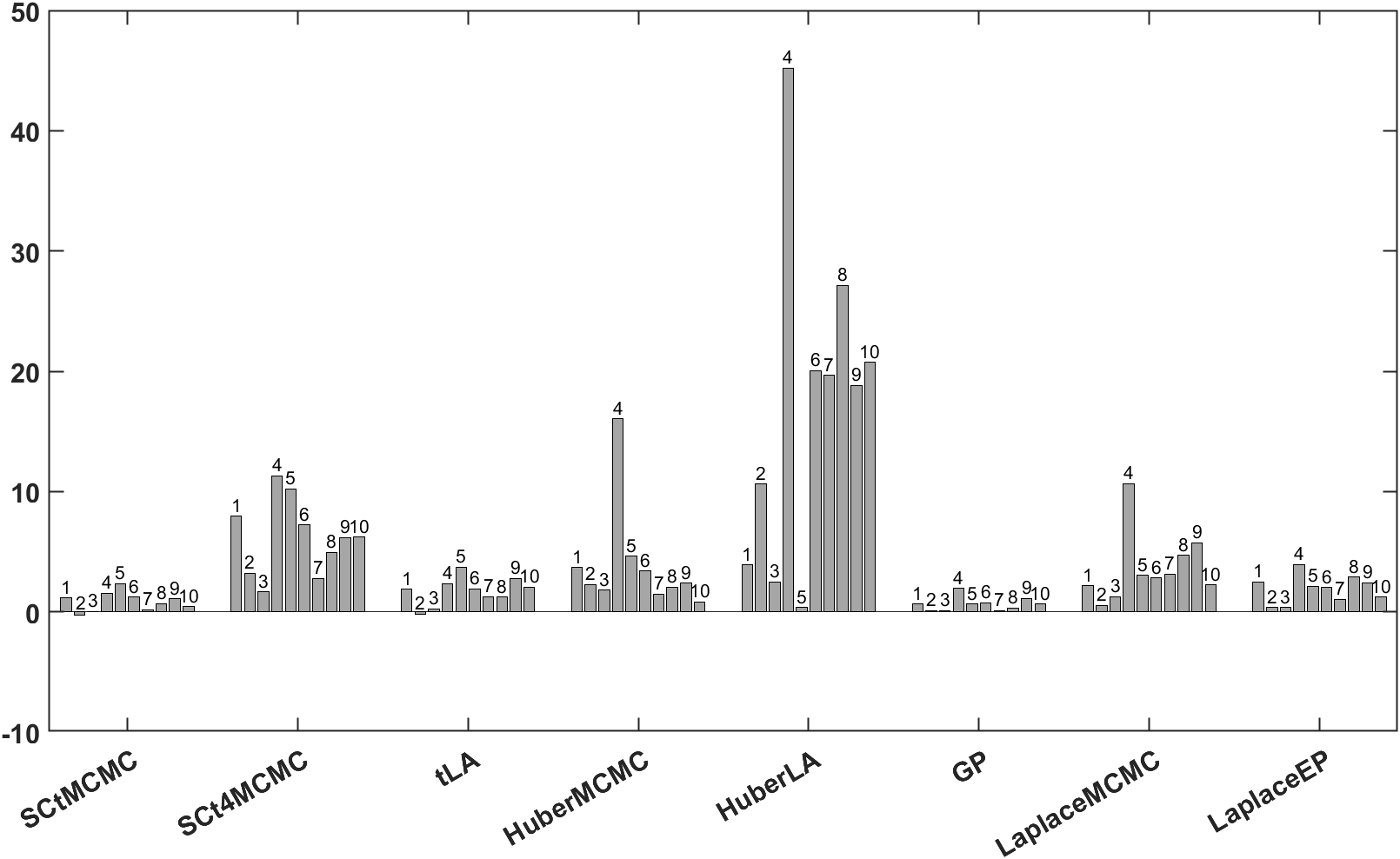} }}%
    \caption{\textit{Indices for the Boston housing data: (a) MAE and (b) NLP.}}%
    \label{fbar1}%
\end{figure}
\begin{figure}
    \centering
    \includegraphics[height=8cm,width=15cm]{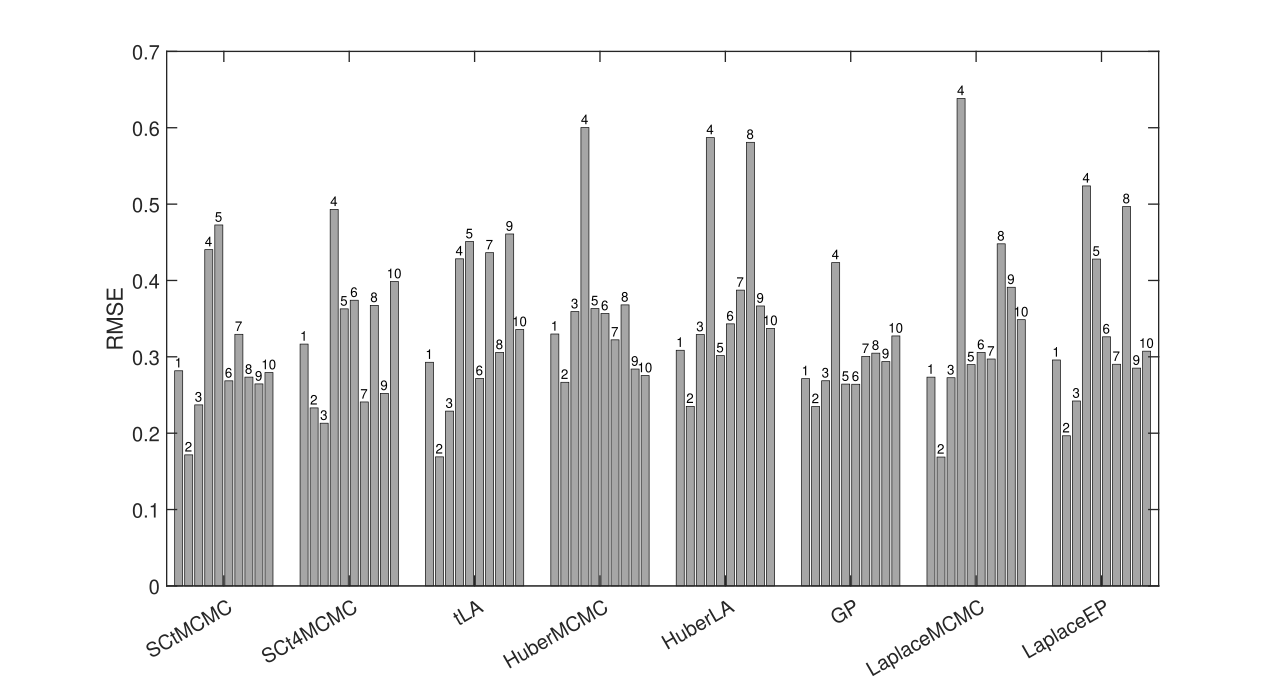}
    \caption{\textit{RMSE values for the Boston housing data.}}
    \label{BS_bar3}
\end{figure}

\begin{figure}
    \centering
    \includegraphics[height=6.7cm,width=14cm]{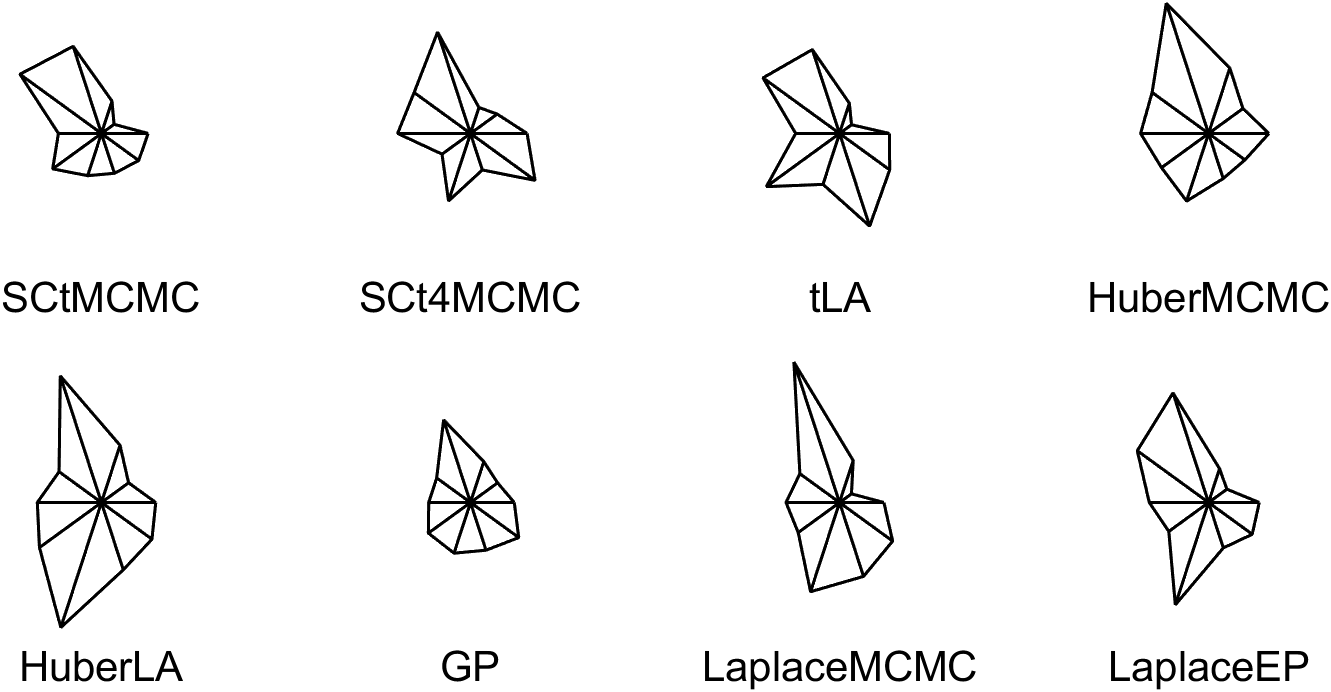}
    \caption{\textit{RMSE folds of the eight considered GP regression models for the Boston housing data.}}
    \label{BS_folds}
\end{figure}

 
The first real dataset we consider is the Boston housing dataset, which was first analyzed by \cite{Harrison1978} and later  popularized by \cite{Kuss2006} and \cite{Jylänki2011} as a reference problem in non-linear regression. The dataset consists of a total of $n=506$ observations, which are split into $10$ folds in order to predict the median price of housing in different parts of the metropolitan Boston area using $13$ features. The MAE and NLP associated with the folds are shown in Figures \ref{fbar1}, where the number on top of each bar indicates the fold. The RMSE of the results obtained from the proposed GP-Huber is compared with those of the aforementioned models in Figure \ref{BS_folds}, where each line corresponds to the RMSE value of each fold. Their values are shown in Figure \ref{BS_bar3}. The performance evaluation values of the results obtained from HuberMCMC are observed to be better than the ones of HuberLA. The results of the GP are noticed to yield the lowest RMSE, MAE, and NLP values. We obtain the results of HuberLA and HuberMCMC by setting the threshold parameter to $2$ to gain efficiency for the normally distributed data. 

\subsection{Case Study 4: Transmission Spectroscopy}
Transmission  spectroscopy  records the  relative  change  in the stellar flux, which is the incident photons per unit area, as a planet travels in front of the star around which it revolves. When the planet faces the star directly, known as a transit, it occludes a fraction of the stellar flux emitted by the star equal to the sky-projected area of the planet as compared to the area of the star, which is referred to as transit depth. The measurement of the total flux over time is known as the light curve. The property on which the transmission spectroscopy relies to estimate the transit curve parameters is the planet’s transit depth, which  dependents on the wavelengths of the transmitted flux. 
For the wavelengths where the planet's atmosphere is opaque due to the absorption of the emitted electromagnetic waves by constituent atoms or molecules, the planet blocks slightly more stellar flux. The variations are measured by binning the light curve into spectrophotometric channels of different wavelengths and by fitting the light curve from each channel separately with a transit model \cite{Kreidberg2018}. 
The sources of error, such as photon noise and instrumental and astrophysical systematics, raise many potential challenges for precise atmosphere characterization. Pointing drift or modifications in the telescope focus influence the spectrum position on the detector to a small degree during transit due to instrumental systematics. Note that instrumental systematics are nothing but what is popularly known as \say{systematic errors} in statistics, which are here  attributed to the atmospheric effects on the physical properties of an instrument. 
The optical state parameters are metered via auxiliary measurements of the spectral trace such as position, width, angle, or other parameters, indicating the state of the detector and optics, which are thought to be the cause of instrumental systematics. Instead of modeling the latter as a linear function of the optical state parameters, \cite{Gibson2011} proposed a non-parametric model by leveraging GPs.  

The observation set obtained from HST- NICMOS includes the light curves for $18$ wavelength channels extracted from n=$638$ spectra along with six optical state parameters, namely the position of the spectral trace along the dispersion axis, $\Delta{X}$, the average position of the spectral trace along the cross-dispersion axis, $\Delta{Y}$, the angle of the spectral trace with the x-axis, $W$, the average width of the spectral trace, $\psi^{s}$, the temperature, $T$, and the orbital phase, $\psi^{H}$. The flux measurements contained in the vector,  $\bm{f}=[f_1,f_2,\hdots,f_n]^{T}$, are recorded at $n$ time instants, $\{t_1,t_2,\hdots,t_n\}$, contained in the time vector, $\bm{t}$, and the optical state parameters are given by $\mathbf{x}_{i}=[\Delta{X}_{i},\Delta{Y}_{i},W_{i},\psi^{H}_{i},T_i,\psi^{s}_{i}]^{T}$ at time instant, $t_i$, collected in the matrix $\mathbf{X}\in \mathbb{R}^{6\times N}$ given by $\mathbf{X}=[\mathbf{x}_1,\hdots,\mathbf{x}_n]$.

\begin{figure}%
    \centering
    \subfloat[\centering  ]{{\includegraphics[width=6.5cm]{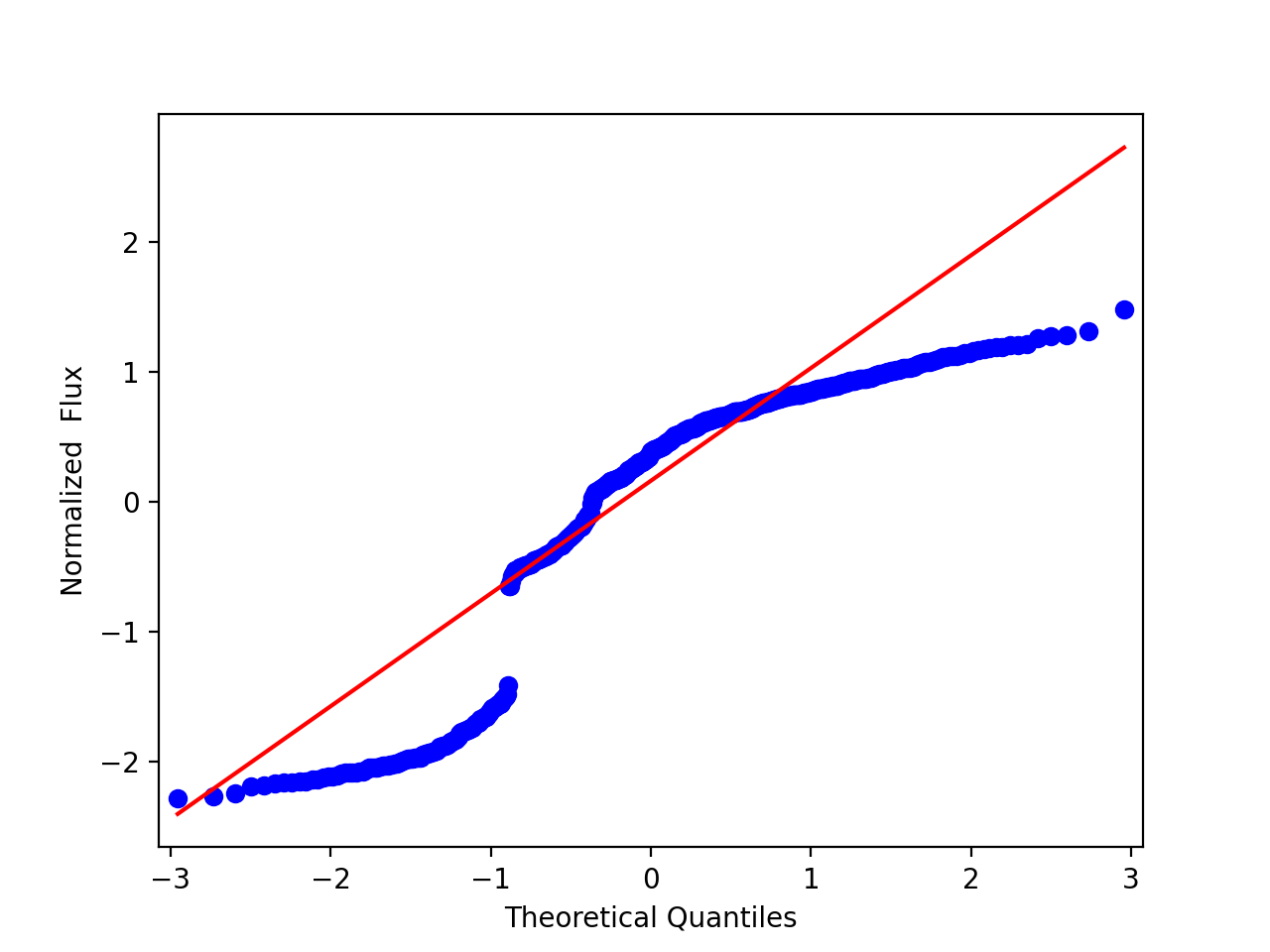} }}%
    \qquad
    \subfloat[\centering ]{{\includegraphics[width=6.5cm]{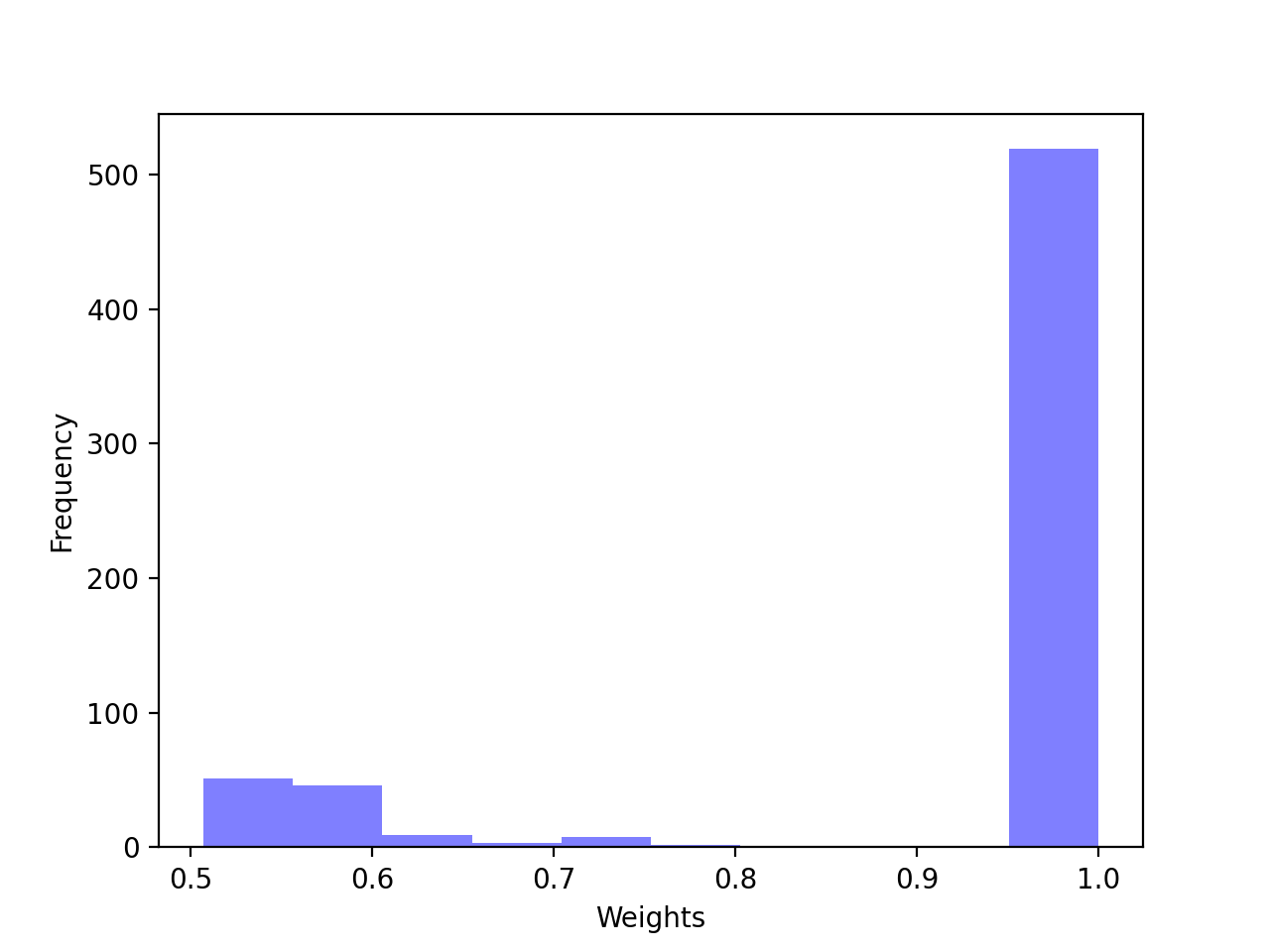} }}%
    \caption{\textit{Plots of the normalized flux and weights corresponding to a single wavelength channel: $(a)$ the QQ plot of the normalized flux measurements and $(b)$ the histogram of the normalized flux and the weights used to downweight the instrumental systematics.}}%
    \label{fdata}%
\end{figure}

In this paper, we extend the work of \cite{Gibson2011} by using the GP-Huber model to estimate the planet-to-star radius ratio, $\rho_{{radius}}$, of HD 189733. As demonstrated earlier, the robustness to outliers of GP-Huber allows us to utilize $517$ measurements associated with four out-of-transit orbits, namely orbit numbers, $\{2,3,4,5\}$, and $137$ measurements associated with one in-transit orbit, namely orbit number $1$. The latter was excluded from the analysis performed by \cite{Gibson2011} as it constitutes much larger systematics effects attributed to the spacecraft settling. The observed transit flux  modeled in the GP framework follows a normal distribution, that is, 
\begin{equation}
    \bm{f}(\bm{t},\mathbf{X})\sim \mathcal{N}(\bm{T}(\bm{t},\bm{\phi}),\mathbf{K}(\mathbf{X},\mathbf{X}|\bm{\theta})).
\end{equation}
where the parameter vector, $\bm{\phi}$, include the parameter of interest, $\rho_{radius}$, and other parameters, namely out-of-transit flux, $f_{oot}$, time gradient, $T_{grad}$,  fixed central transit time, $T_{0}$, orbital period, $P$, limb darkening coefficient, $c_{1}$, limb darkening coefficient, $c_{2}$.  The transit vector function, $\bm{T}(\bm{t},\bm{\phi})$, is hereafter referred to as mean function parameter vector. The non-variable mean function parameters are fixed or calculated as stated in \cite{Gibson2011}. Along with the planet-to-star radius ratio, the other mean function parameters are the parameters of a linear baseline model, $f_{oot}$ and $T_{grad}$. The covariance matrix, $\mathbf{\Sigma}(\mathbf{x}_{i},\mathbf{x}_{j}|\bm{\theta})$, is the covariance between two output flux measurements defined as a function of the distance between optical state parameters, $(\mathbf{x}_{i},\mathbf{x}_{j})$, given by 
\begin{equation}\label{eq045}
  {K}_{ij} = k(\mathbf{x}_{i},\mathbf{x}_{j})+\delta_{ij}\sigma^{2},
\end{equation}
where $k(\cdot,\cdot)$ is a Gaussian kernel discussed in Section \ref{sec:GPE}. We consider the analytical quadratic limb darkening transit function proposed by \cite{Mandel2002}. To develop the model in the GP-Huber framework, we begin with the computation of the projection statistics given by \eqref{eq36} on optical state parameter values contained in $\mathbf{X}$, which are separately for each orbit. Figure \ref{fdata}(a) plots the normalized flux measurements vs. theoretical quantiles of the normal distribution. We observe in Figure \ref{fdata}(b) that the weights of the errors of the measurements associated with the orbits are centered around $0.55$ and those of the remaining measurements associated with the orbits are nearly equal to $1$. The threshold parameter, $b$, is set to $1.5$ to achieve good robustness and efficiency at data distributed normally. Analogous with \eqref{eq044}, we assume that the observed transit flux vector, $\bm{f}=\bm{f}(\bm{t},\mathbf{X})$, in the GP-Huber framework follows a normal distribution, that is,  
\begin{equation}
    \bm{f}|\bm{T}(\bm{t},\mathbf{X}),\mathbf{X},\bm{\phi},\bm{\theta},\bm{\sigma}^{2}) \sim \mathcal{N}\left(\bm{T}(\bm{t},\mathbf{X}),\mathbf{\Sigma}+\mathbf{K}\right).
\end{equation}
Here, the variances of the parameter errors are included individually for each observation data point as diagonal elements of the covariance matrix, $\mathbf{\Sigma}$. Therefore, along with the transit function parameter vector, $\bm{\phi}$, and the kernel function parameter vector, $\bm{\theta}$, we need to infer the variance vector,  $\bm{\sigma}^{2}=[\sigma^{2}_{g}, \bm{\sigma}^{2}_{l}]^T$, containing the variance, $\sigma^{2}_{g}$, of the inlying observation data points and the variances contained in $\bm{\sigma}^{2}_{l}$ of the outlying observation data points. As per the scale mixture representation of the Laplace error distribution, the variance vector, $\bm{\sigma}^{2}$, follows an exponential distribution with rate parameter vector, $\bm{\beta}=[\beta_{g}, {\beta}_{l}]^T$. The kernelized covariances are represented by $\mathbf{K}$ with entries $K_{ij}=k(\mathbf{x}_{i},\mathbf{x}_{j})$.

The joint un-normalized log-posterior function of $\bm{\phi}$, $\bm{\beta}$, and  $\bm{\theta}$ with the gamma prior probability density function, $p(
\bm{\theta})=\frac{1}{\bm{l}}\textrm{exp}\left(\frac{-\bm{\theta}}{\bm{l}}\right)$, over the covariance function hyperparameters is given by
\begin{multline}\label{eq51}
     \textrm{log\;}{P}(\bm{\phi},\bm{\theta},\bm{\sigma}^{2},\bm{\beta}|\bm{f},\mathbf{X},\bm{\zeta}) = \textrm{log\;}\left(\mathcal{L}(\mathbf{r}|\mathbf{X},\bm{\phi},\bm{\theta},\bm{\sigma}^{2})
  \right)-\frac{\tau}{l_{\tau}}-\sum_{i=1}^{d}\left(\frac{1}{s_i l_{i}}\right)\\
  +\textrm{log}(\bm{\beta})-\bm{\beta}^{T}\bm{\sigma}^{2}+\textrm{log}(p(\bm{\beta}|\bm{\zeta}))+\textrm{C}, 
\end{multline}
where $l_{\tau}$ is a parameter of the gamma prior associated with hyperparameter $\tau$ and the  C represents additional constant terms. The samples of ${\beta}_{l}$ are generated from log uniform distribution to lay a non-informative prior with parameter vector, $\bm{\zeta}$, whereas $p(\beta_{g})$ is a degenerate probability density function. 

The challenging task now is to infer the parameter, $\rho_{radius}$, from the joint posterior distribution of $(\bm{\phi},\bm{\theta},\bm{\sigma}^{2},\bm{\beta})$ given by \eqref{eq51} in which the log-likelihood term is expressed as
\begin{equation}
    \textrm{log}\;\mathcal{L}(\mathbf{r}|\mathbf{X},\bm{\phi},\bm{\theta},\bm{\sigma}^{2})= \frac{-1}{2}\mathbf{r}^{T}(\mathbf{\Sigma}+\mathbf{K})^{-1}\mathbf{r}-\frac{1}{2}\textrm{log}|\mathbf{\Sigma}+\mathbf{K}|-\frac{n}{2}\textrm{log}(2\pi)+\textrm{log}(1-\varepsilon),
\end{equation}
where $\mathbf{r}=\bm{f}-\bm{T}(\bm{t},\mathbf{X})$. One of the approaches is to use the Bayesian method that seeks the posterior distribution of $\rho_{radius}$ by marginalizing over the other parameters of the mean function parameters $\bm{\phi}$ and the covariance function hyperparameters, $\bm{\theta}$ using MCMC methods.
The other method proposed as the type-II maximum likelihood method by \cite{Gibson2011}, where the hyperparameters, $\bm{\theta}$ and $\bm{\sigma}^{2}$ and the variable mean function parameters, $f_{oot}$ and $T_{grad}$ are fixed at their maximum a posteriori estimates. Formally, we have 
\begin{equation}
    (\hat{\bm{\phi}},\hat{\bm{\theta}},\hat{\bm{\sigma}}^{2},\hat{\bm{\beta}})= \underset{\bm{\phi},\bm{\theta},\bm{\sigma}^{2},\bm{\beta}}{\textrm{arg}\; \textrm{max}}\; \textrm{log}\;P(\bm{\phi},\bm{\theta},\bm{\sigma}^{2},\bm{\beta}|\bm{f},\mathbf{X},\bm{\zeta}).
\end{equation}
And the posterior distribution of the parameter of interest $\rho_{radius}$ is obtained by marginalizing the joint posterior distribution $p(\bm{\phi},\bm{\theta},\bm{\sigma}^{2},\bm{\beta})$  over the hyperparameters and the rest of the mean function parameters. In the standard type II maximum likelihood method, the hyperparameters are fixed to their maximum likelihood estimates i.e. by maximizing the evidence $p(\mathcal{D}|\bm{\phi},\bm{\theta},\bm{\sigma}^{2})$. 

\begin{figure}[!t]%
    \centering
    \subfloat[\centering ]{{\includegraphics[height=5.0cm,width=6.5cm]{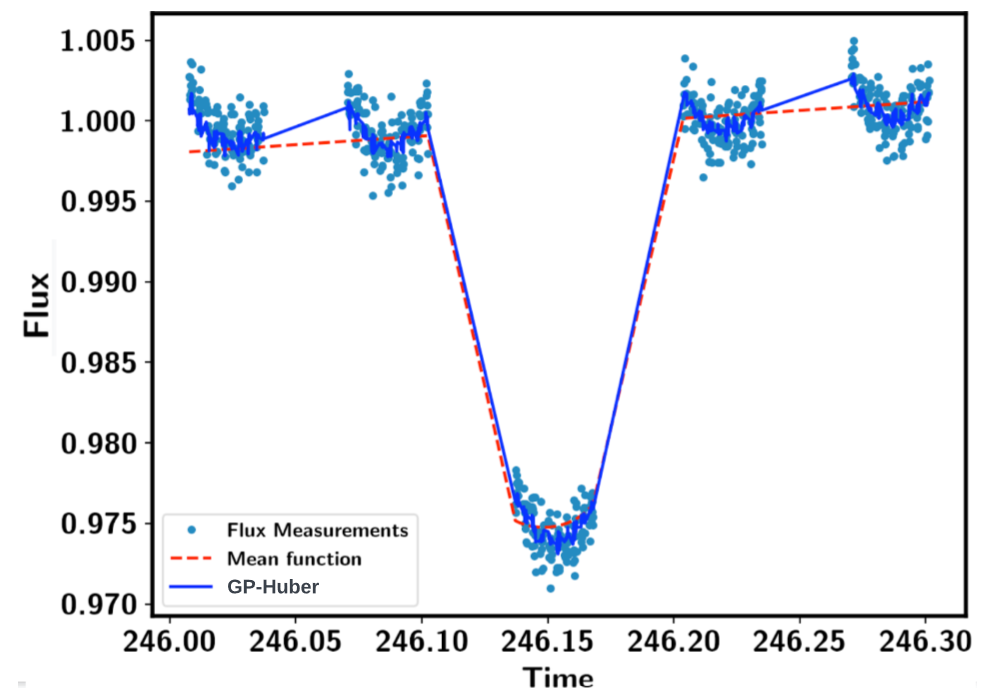} }}%
    \qquad
    \subfloat[\centering  ]{{\includegraphics[height=5.0cm,width=6.5cm]{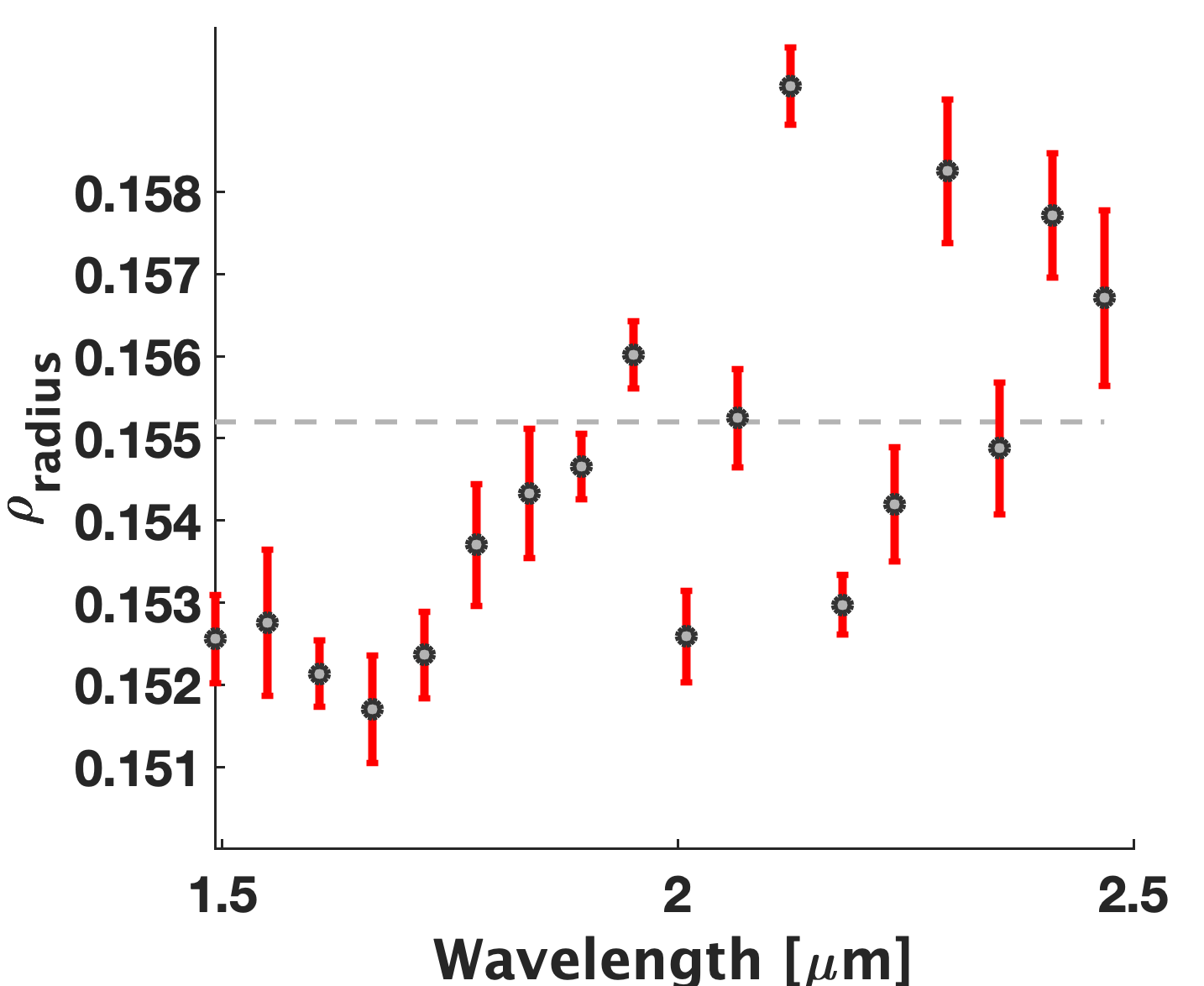} }}%
    \caption{\textit{Transit curve fit and estimated $\rho_{radius}$. $(a)$ Transit curve mean function $T(t,\bm{\theta})$ and GP-Huber model fit; $(b)$ results of planet-to-star radius ratios ($\rho_{radius}$) obtained from GP-Huber with error-bars.}}%
    \label{radius}%
\end{figure}

Figure \ref{radius}(a) shows the transit fit obtained for one wavelength channel. We notice that the fit associated with all the orbits fits the central data points and excludes the outlying data points. 
Figure \ref{radius}(b) shows the estimated $\rho_{radius}$ obtained using MCMC integration over the rest of the mean function parameters $\bm{\phi}$ and hyperparameters $\bm{\theta}$ along with the values estimated from the white light curve represented as the white dashed line. The results of the planet-to-star radius ratio for each wavelength obtained from the GP-Huber model are shown in Table \ref{tabrho} along with those obtained from the model described in \cite{Gibson2011}, hereafter referred to as Gibson2012, where $\Delta\rho_{radius}$ represents the estimated uncertainty. Most of our results agree with the results obtained from the Gibson model except for wavelength channels $1.665 \mu$m and $2.124 \mu$m. 

\section{CONCLUSION}
\label{sec:conc}
In this paper, we developed a robust Gaussian process regression with Huber density (GP-Huber) where the residuals are standardized based on robust distances called projection statistics to identify and downweight the bad leverage points and vertical outliers. We develop the Laplace approximation (HuberLA) and hybrid  Monte Carlo method (HuberMCMC) for approximate Bayesian inference.
The proposed GP-Huber model is particularly promising for a wide range of thick-tailed noise distributions and normal noise distribution. It does not require any additional likelihood parameters specific to the noise distributions. We demonstrate the performance of the GP-Huber on two numerical and two real-world datasets. Firstly, we show the applicability of the proposed model on the Neal dataset where the added error follows the normal, the Student's t, the Laplace, and the Cauchy distribution. Secondly, we evaluate the performance of the GP-Huber on the Friedman dataset using measures RMSE, MAE, and NLP. Thirdly, we consider our first demonstration on a real-world dataset called Boston housing data where the task is to predict the median housing prices of the households in the Boston area. Fourthly, we predict the planet-to-star radius of the planetary system HD$189733$ using the recorded flux obtained from the HST-NICMOS instrument. Our simulation results on the first two numerical datasets show that the GP-Huber with Laplace approximation yields better performance than the GP-Huber with the MCMC method. For the Boston housing dataset, the HuberMCMC method gives better results. To solve the non-linear regression problem in inferring the planet-to-star radius ratio, $\rho_{radius}$, using the MCMC method, we marginalize the joint posterior density of parameters and hyperparameters of the transit model for the quantity of interest, $\rho_{radius}$. In future work, we will assess the performance of the GP-Huber model under other thick-tail distributions, such as the $\alpha$ stable distributions. We will also investigate the use of high breakdown estimators to achieve a high breakdown point for highly corrupted real-world training datasets. 

\begin{table}[!t]
    \centering
      \caption{Results of the planet-to-star radius ratio obtained from Gibson$2012$ and GP-Huber.}
    \begin{tabular}{ccccc}
    \hline
    {Wavelength} & \multicolumn{2}{c}{Results from model in Gibson$2012$} & \multicolumn{2}{c}{Results obtained from GP-Huber}\\
       ($\mu$m) & $\rho_{radius}$& $\Delta\rho_{radius}$ & $\rho_{radius}$& $\Delta\rho_{radius}$ \\
      \hline
      2.468&0.15545&0.00077&0.15525&0.00071\\
2.411&0.15520&0.00052&0.15771&0.0008911\\
2.353&0.15455&0.00044&0.15488&0.0004021\\
2.296&0.15513&0.00057&0.15825&0.0006526\\
2.238&0.15512&0.00041&0.1542&0.0005276\\
2.181&0.15504&0.00051&0.15297&0.0007462\\
2.124&0.15417&0.00066&0.15928&0.0007869\\
2.066&0.15508&0.00066&0.15525&0.000399\\
2.009&0.15393&0.00036&0.15259&0.0004077\\
1.951&0.15595&0.00051&0.15602&0.0005586\\
1.894&0.15549&0.0006&0.15466&0.0005988\\
1.837&0.15513&0.00053&0.15433&0.0004704\\
1.779&0.15534&0.00051&0.1537&0.0003601\\
1.722&0.15447&0.00087&0.14937&0.0006938\\
1.665&0.15429&0.00064&0.1517&0.000871\\
1.607&0.15266&0.00062&0.15213&0.0008045\\
1.55&0.15359&0.00073&0.15276&0.0007583\\
1.492&0.15367&0.00118&0.15256&0.0010653\\
    \end{tabular}
    \label{tabrho}
\end{table}

\begin{funding}
 The authors gratefully acknowledge the financial support of NSF via grant ID 1917308.
\end{funding}

\begin{supplement}
\stitle{Supplement A}
\sdescription{Python code that was used for the estimated results for eighteen wavelengths from Section 4.2 along with the package GeePea and the dataset. (.tar file)}
\stitle{Supplement A}
\sdescription{Matlab code used for the demonstration of the GP-Huber model on the Gaussian, the Laplace, and the Student's t- distributions in Section 4.1. (.tar file)}
\end{supplement}




\bibliographystyle{apa}

\end{document}